\documentclass[10pt]{article}

\usepackage{amsmath}

\usepackage{array}

\usepackage{appendix}

\usepackage{tocloft}                   

\usepackage{graphicx}

\usepackage{amsfonts}

\usepackage{amssymb}

\usepackage{mathrsfs}

\usepackage{yfonts}

\usepackage{euscript}

\usepackage{centernot}                 

\usepackage{ifsym}                     

\usepackage{upgreek}

\usepackage{mathtools}

\usepackage{color}

\usepackage{slantsc}
\usepackage{calligra}

\usepackage{bbold}          

\usepackage[T1]{fontenc}

\usepackage{epsf}

\usepackage{latexsym}

\usepackage{tipa}

\usepackage{makeidx}

\makeindex

\def\b{\begin{equation}}

\def\e{\begin{equation}}

\def\be{\begin{equation}}              

\def\ee{\end{equation}}

\def\beq{\begin{equation}}

\def\eeq{\end{equation}}

\def\bea{\begin{eqnarray}}

\def\eea{\end{eqnarray}}

\def\m{\mbox{ }}

\def\mma {\m , \m \m }

\def\!{\hspace{-1.6667em}}

\def\n{\noindent}

\def\u{\underline}

\def\s{\stackrel}

\def\slTheta{\mathit{\Theta}}                     

\def\biN{\mbox{\boldmath$N$}}

\def\mA{\mbox{A}}

\def\mC{\mbox{C}}                        

\def\mI{\mbox{I}}                        

\def\mL{\mbox{L}}

\def\mO{\mbox{O}}

\def\mP{\mbox{P}}

\def\mS{\mbox{S}}                        

\def\mW{\mbox{W}}

\def\mo{\mbox{o}}

\def\mp{\mbox{p}}

\def\muu{\mbox{u}}

\def\bupSigma{\mbox{\boldmath$\Sigma$}}                 

\def\sg{\mbox{\scriptsize g}}

\def\so{\mbox{\scriptsize o}}

\def\st{\mbox{\scriptsize t}}

\def\sA{\mbox{\scriptsize A}}

\def\sL{\mbox{\scriptsize L}}

\def\sR{\mbox{\scriptsize R}}

\def\sS{\mbox{\scriptsize S}}

\def\sumi2{\sum\mbox{}_{\mbox{}_{\mbox{\scriptsize $i$=1}}}^2}

\def\sumi3{\sum\mbox{}_{\mbox{}_{\mbox{\scriptsize $i$=1}}}^3}

\def\sumABcycles3{\sum\mbox{}_{\mbox{}_{\mbox{\scriptsize cycles $A,B$=1}}}^{3}}

\def\sumCDcycles3{\sum\mbox{}_{\mbox{}_{\mbox{\scriptsize cycles $C,D$=1}}}^{3}}

\def\sumj3{\sum\mbox{}_{\mbox{}_{\mbox{\scriptsize $j$=1}}}^3}

\def\sumk3{\sum\mbox{}_{\mbox{}_{\mbox{\scriptsize $k$=1}}}^3}

\def\prodiA1{\prod\mbox{}_{\mbox{}_{\mbox{\scriptsize $i$=1}}}^{A - 1}}

\def\bigtimes{\mbox{\Large $\times$}}

\def\pa{\partial}                                                   

\def\es{\m = \m}

\def\:={\m := \m}

\def\=:{\m =: \m}

\def\FrA{\mbox{$\mathfrak{A}$}}                                

\def\FrT{\mathfrak{T}}                                         

\def\FrC{\mbox{$\mathfrak{C}$}}                                
                                                       
\def\sFrR{\mbox{\scriptsize $\mathfrak{R}$}}                   
\def\tFrR{\mbox{$\mathfrak{f}$}}                               

\def\FrS{\mbox{\Large $\mathfrak{s}$}}                         
\def\tFrS{\mbox{\footnotesize$\mathfrak{s}$}} 
\def\sFrM{\mbox{\scriptsize$\mathfrak{M}$}}                    

\def\lFrg{\mbox{\Large$\mathfrak{g}$}}                         
\def\FrT{\mbox{\boldmath$\mathfrak{T}$}}                       
\def\sFrT{\mbox{\scriptsize$\mathfrak{T}$}}                    

\def\Hilb{\mbox{{\boldmath$\mathfrak{H}$}ilb}}                 
\def\bFrB{\mbox{\boldmath$\mathfrak{B}$}}                      

\def\FrQ{\mbox{\Large $\mathfrak{q}$}}                               
\def\bFrC{\mbox{\boldmath$\mathfrak{C}$}}                            
\def\bFrL{\mbox{\boldmath$\mathfrak{L}$}}                            

\def\Phase{\mbox{{\boldmath$\mathfrak{P}$}hase}}                     

\def\bFrR{\mbox{\boldmath$\mathfrak{R}$}}                            
\def\Rig-Phase{\bFrR\mbox{ig-}\Phase}                                
                                                                													   
\def\FrR{\mbox{\boldmath$\mathfrak{R}$}}                             

\def\sFrR{\mbox{\scriptsize\boldmath$\mathfrak{R}$}}                 
\def\tFrR{\mbox{\tiny\boldmath$\mathfrak{R}$}}                       

\def\bFrU{\mbox{\boldmath$\mathfrak{U}$}}                            
					
\def\bFrJ{\mbox{\boldmath$\mathfrak{J}$}}                            

\def\bFrR{\mbox{\boldmath$\mathfrak{R}$}}                            

\def\bFrI{\mbox{\boldmath$\mathfrak{I}$}}                            
					
\def\bFrH{\mbox{\boldmath$\mathfrak{R}$}}                            
	
\def\bFrR{\mbox{\boldmath$\mathfrak{R}$}}                            

\def\1mat{\u{\u{1}}}                                                 

\def\Leib{\bFrL\mbox{eib}}                                           

\def\Sym{\bFrS\mbox{ym}}                                             
\def\Uni{\bFrU\mbox{ni}}                                             

\def\Lag{\bFrL\mbox{ag}}                                             

\def\Jac{\bFrJ\mbox{ac}}                                             
\def\Incip{\bFrI\mbox{ncip}}                                       

\def\Reg{\bFrR\mbox{eg}}                                             
                                                                     
\def\Co{\bFrC\mbox{o}}                                               
  
\def\Hopf{\bFrH\mbox{opf}}                                           

\def\Positive-Modespace{\mbox{{\boldmath$\mathfrak{M}$}odespace$^+$}}

\def\POSITIVE-MODESPACE{\mbox{{\boldmath$\mathfrak{M}$}ODESPACE$^+$}}
                                                                                                                             														
\def\bFrS{\mbox{\Large $\mathfrak{s}$}}                              
			
\def\bFrG{\mbox{ $\mathfrak{G}$}}                                    %
\def\Riem{\bFrR\mbox{iem}}                                           
\def\Superspace{\bFrS\mbox{uperspace}}                               

\def\FrO{\mbox{$\mathfrak{O}$}}                                      

\def\Top{\FrT\mo\mp}

\def\Big{\bFrB\mbox{ig}}

\def\Grand{\bFrG\mbox{rand}}

\def\sFrC{\mbox{\boldmath\scriptsize$\mathfrak{C}$}}                        

\def\Kin-Hilb{\mbox{{\boldmath$\mathfrak{K}$}in-\Hilb}}                     

\def\Mid-Hilb{\mbox{{\boldmath$\mathfrak{M}$}id-\Hilb}}                     

\def\Dyn-Hilb{\mbox{{\boldmath$\mathfrak{D}$}yn-\Hilb}}                     

\def\5Star{\mbox{\Large$\star$}}              

\begin{document}

\begin{titlepage}

\begin{center}

\Large{\bf BACKGROUND INDEPENDENCE:} \normalsize

\vspace{0.1in}

\large{\bf $\mathbb{S}^1$ and $\mathbb{R}$ absolute spaces differ greatly in Shape-and-Scale Theory}

\vspace{0.2in}

{\normalsize \bf Edward Anderson$^*$}

\vspace{.2in}

\end{center}

\begin{abstract}

\n Kendall-type Shape(-and-Scale) Theory on $\mathbb{R}^d$ involves $N$ point configurations therein quotiented by some geometrically meaningful automorphism group. 
This occurs in Shape Statistics, the Classical and Quantum $N$-body Problem and as a model for many aspects of Generally Relativistic theories' Background Independence.

Shape-and-Scale theory on $\mathbb{S}^1$ is significant at the level of `rubber shapes' as 1 of only 3 classes of connected-without-boundary absolute spaces.
It is also the first torus $\mathbb{T}^d$ and real-projective space $\mathbb{RP}^d$ as well as the first sphere $\mathbb{S}^d$, 
%
%
and in fact behaves far more like $\mathbb{T}^d$ than $\mathbb{S}^d$ for $d \geq 2$.
We now investigate the $\mathbb{S}^1$ case at the geometrical level.  
With $Isom(\mathbb{S}^1) = SO(2)$ itself a $\mathbb{S}^1$, the shape-and-scale $N$-body configuration spaces are systematically $\mathbb{T}^{N - 1}$.    
We show moreover that 3 points on the circle $\mathbb{S}^1$ already suffices for major differences to occur relative to on the line $\mathbb{R}$.
Scale is now obligatory. 
Totally antipodal configurations are as significant as the maximal collision.
Topologically, partially antipodal configurations play an equivalent role to right angles: specifically a $d \geq 2$ notion on $\mathbb{R}^d$.    
Using up less and more arc than an antipodal configuration are the respective topological analogues of acute and obtuse triangles.    
The idea that quotienting out geometrical automorphisms banishes an incipient notion of absolute space is dead.  
Such indirect modelling is, rather, well capable of remembering the incipient absolute space's topology.
Thus topological considerations of Background Independence have become indispensible even in mechanics models. 
In General Relativity, this corresponds to passing from Wheeler's Superspace to Fischer's Big Superspace.  

\end{abstract}

\n {\bf PACS}: 04.20.Cv. 

\m 

\n {\bf Physics keywords}: Background Independence, Absolute versus Relative Motion Debate, configuration spaces, dynamical and quantization aspects of General Relativity.

\m

\n {\bf Mathematics keywords}: Shape Theory, Applied Topology, Shape Statistics.

\vspace{0.1in}
  
\n $^*$ Dr.E.Anderson.Maths.Physics@protonmail.com

\vspace{0.1in}

\end{titlepage}

\section{Introduction}\label{Intro-I}

\n Shape Theory in David Kendall's sense \cite{Kendall84, Kendall89, Small, Kendall, FileR, Bhatta, I, II, III, QuadI, IV, PE16} 
-- of $N$-point configurations (constellations) in $\mathbb{R}^d$ quotiented by similarity transformations -- 
has more recently been shown to admit a wide variety of generalizations.

\m 

\n{\bf Generalization 1} Quotienting out by other groups, such as Euclidean \cite{M02, M05, Cones, FileR}, 
                                                  affine \cite{Sparr-GT09, Bhatta, AMech, PE16, Affine-Shapes}, 
										      projective \cite{MP05-KKH16, Bhatta, PE16} 
											   conformal \cite{AMech} 
									  and supersymmetric \cite{AMech} groups.   

\m 

\n{\bf Generalization 2} Considering other carrier spaces (alias absolute spaces in the physically realized case) in place of $\mathbb{R}^d$, 
such as spheres         $\mathbb{S}^d$  \cite{Kendall87, FileR, ASphe}, 
tori                    $\mathbb{T}^d$  \cite{JM00, ATorus},
real projective spaces  $\mathbb{RP}^d$ \cite{MP05-KKH16, Bhatta, PE16}, 
generic spaces                          \cite{Generic}, 
Minkowski spacetime     $\mathbb{M}^d$  \cite{Project-1}
and a local version of Shape Theory     \cite{Local-Shapes}.  

\m 

\n Such studies are rooted in the topology and geometry of the corresponding spaces of shapes. 
Some applications are as follows. 

\m 

\n{\bf Application 0} Shape Theory has been shown to be capable of yielding interesting geometrical results in their own right 
\cite{Kendall84, Kendall89, Small, Kendall, MIT, III, A-Pillow, IV}.    
 
\m

\n{\bf Application 1} The main application to date has moreover been to Shape Statistics (see \cite{Small, Kendall, JM00, Bhatta, DM16, PE16} for reviews), 
itself very widely applicable throughout the STEM subjects: from Biology and Archaeology to Astrophysics, Robotics, Image Analysis and Computer Vision. 

\m 

\n{\bf Application 2} See furthermore e.g.\ \cite{LR95-97, M02, M05, AF, +Tri, FileR, QuadI, APoT1, AKendall, ASphe, AConfig, ABook, I, II, III, A-Monopoles} 
for Shape Theory and Shape-and-Scale Theory (jointly Relational Theory)
now also beeing established as part of Theoretical Mechanics \cite{BB82, LR95-97, M02, M05, FORD, Cones, FileR, QuadI, AMech, ABook}.  
This is of relevance to the Absolute versus Relational Motion Debate -- which dates back at least as far as Leibniz versus Newton \cite{Newton, L, M, DoD-Buckets, ABook, Generic} --
and has additionally now been treated at the quantum level                                                                       \cite{AF, +Tri, FileR, QuadI, ABook, Forth}.

\m 

\n{\bf Application 3} Relational Theory has furthermore been shown to model \cite{B94I, KieferBook, FileR, AKendall, ABook} 
many aspects of Classical and Quantum General Relativity's Background Independence \cite{A6467, K92-I93, Giu06, BI, APoT3, ABook, Project-1, Generic} 
with its ties to the structure of GR configuration spaces                          \cite{Battelle, DeWitt67, DeWitt, Fischer, FM96, Giu09, ABook} 
and to the Problem of Time                                                         \cite{Battelle, DeWitt67, K92-I93, APoT1, FileR, APoT3, ABook}.

\m 

\n As regards establishing a broader and deeper foundational basis for Relational Theory, firstly the {\it Relational Aufbau Principle} \cite{I} -- 
to start with small $N$, $d$ and automorphism group being quotiented $\lFrg$ and build up -- has proven to be very useful. 
This is because smaller $d$, $N$ and $\lFrg$ re-enter the study of larger such, in particular as submanifolds, strata and significant subgroups.  
\n Secondly, considering the less structured `rubber shapes' -- Topological Relational Theory \cite{I, II, III, Top-Shapes, IV} -- 
provides just three coarser universality classes for carrier spaces which are connected manifolds without boundary, 
features of which are then reflected in whichever more geometrically structured theory one builds thereupon. 
These three classes are $d \geq 2$ manifolds as a single class, the real line $\mathbb{R}$, and the current article's circle $\mathbb{S}^1$.  

\m 

\n Relational Theory on $\mathbb{R}^1$ having been extensively studied elsewhere \cite{JM00, AF, Cones, FileR, I, II, Top-Shapes}, the subject of the current article -- 
   Relational Theory on $\mathbb{S}^1$ -- appears as somewhat of a gap in the rubber shape universality classes and lowest levels of the Relational Aufbau Principle. 
With \cite{Top-Shapes} covering Topological Relational Theory of $\mathbb{S}^1$, 
what is left for us to do is give a detailed metric-level account of Relational Theory on $\mathbb{S}^1$.

\m 

\n We consider $\mathbb{S}^1$'s general constellation spaces, isometry group -- $SO(2) = U(1)$ -- and subsequent quotient relational spaces in Sec 2. 
Scale is now obligatory, and the shape-and-scale spaces are tori $\mathbb{T}^{N - 1}$.   
We also outline discrete quotients of relational spaces including the maximally-quotiented case: Leibniz space \cite{Kendall89, Kendall, FileR, QuadI, ABook, I}.  
We build up to the minimal relationally nontrivial unit of $N = 3$ points on $\mathbb{S}^1$ in Secs 3 to 5. 
We show that the Leibniz space for $N = 3$ for relational theory on $\mathbb{S}^1$ realizes the `30$^{\so}$-60$^{\so}$-90$^{\so}$ set square' triangle.  
We give further levels of structure thereupon in Sec 6.  
The discrete quotients turn $\mathbb{S}^1$'s relational space tori into manifold chunks, some with boundary and/or conical singularities.   
For equal particle masses, the discrete-quotient chunks are (modulo topological identification) 
built from identical tiles \cite{AF, +Tri, FileR, I, II, III, IV} in the shape of the Leibniz space. 

\m 

\n The $N = 3$ Shape-and-Scale Theory on $\mathbb{S}^1$ moreover already suffices for major differences to occur relative to its $\mathbb{R}$ counterpart.
Totally antipodal configurations are as significant as the maximal collision.
We show that partially antipodal configurations play a relational-space-topologically-equivalent role to right angles for triangles in 2-$d$.  
Due to such qualitative differences with relational theory on $\mathbb{R}$, 
the idea that quotienting out geometrical automorphisms banishes an incipient notion of absolute space is dead.  
Such indirect modelling is, rather, well capable of remembering the incipient absolute space's topology.  
C.f.\ QM's topological sensitivity \cite{I84}, though the current article's main point is that Background Independence is already topologically sensitive at the {\sl classical} level. 
Topological considerations of Background Independence have thus become indispensible even in mechanics models. 
In GR theories, this is handled by passing from Wheeler's Superspace \cite{Battelle, DeWitt67, DeWitt, Fischer, FM96, Giu09} 
to Fischer's far more complicated notion of Big Superspace \cite{Fischer}.  

\m

\n We give a Dynamics application in Sec 8 -- boundary conditions for free motion on Leibniz space -- and a Probaility application in Sec 9: 
the counterpart of Lewis Carroll's pillow problem \cite{Carroll, Guy-Portnoy} of what Prob(obtuse) is for triangles,  
for which Shape Theory provides both an answer \cite{Small, MIT, A-Pillow} and quite a lot of calculable variants \cite{A-Pillow}. 
We conclude in Sec 10, including firstly brief comparison with $\mathbb{T}^d$, $\mathbb{S}^d$ and $\mathbb{RP}^d$ Shape-and-Scale Theories 
since $\mathbb{S}^1$ is the first of the $\mathbb{T}^d$ and of the $\mathbb{RP}^d$ Relational Theories as well as of the $\mathbb{S}^d$ ones.  
Secondly, the above GR point is further developed in the last part of the Conclusion.

\section{$\mathbb{S}^1$ carrier space, constellation space, automorphisms, relational space}

\n{\bf Definition 1} The {\it carrier space} $\FrC^d$, alias absolute space in the physically realized case,  
is an at-least-provisional model for the structure of space.

\m 

\n{\bf Example 1} The most usually considered carrier space is $\mathbb{R}^d$.

\m 

\n{\bf Example 2} the current article's carrier space is however   
\be
\FrC^d = \mathbb{S}^1                                                                                           \m . 
\ee 
\n{\bf Definition 2} A {\it constellation} is a collection of $N$ points on a given carrier space $\FrC^d$. 
The {\it constellation space} $\FrQ(N, \FrC^d)$ is the collection of all possible constellations. 

\m 

\n{\bf Remark 1} In some physical applications, the points model material particles (classical, and taken to be of negligible extent).
Because of this, we subsequently refer to constellations as consisting of points-or-particles. 

\m 

\n{\bf Remark 2} Constellations are allowed to include coincident points, or collisions of particles; 
we subsequently refer to these special configurations as coincidences-or-collisions. 

\m 

\n The current article's constellation spaces are thus, for point-or-particle number $N$,   
\be
\FrQ(N, \mathbb{S}^1) :=  \bigtimes_{i = 1}^N  \mathbb{S}^1         
                       =  \mathbb{T}^N                       \m : \m \mbox{ the $N$-dimensional torus }                  \m . 
\ee 
\n{\bf Remark 3} Shape(-and-Scale) Theory furthermore takes some group $\lFrg$ of automorphisms of $\FrC^d$ -- or $\FrQ$ by its product group structure -- 
and regards these as irrelevant to the modelling in question. 
This includes e.g. eliminating the Euclidean group of translations and rotations in a bid to free one's modelling from absolute space. 
This is an example of quotienting out an isometry group. 
The $\mathbb{S}^1$ analogue of this is to quotient out 
\be 
\lFrg = Isom(\mathbb{S}^1) = SO(2) = U(1) = \mathbb{S}^1                                    \m \mbox{ as a manifold}            \m . 
\ee
Clearly 
\be 
\mbox{dim}(\lFrg)  =  1                                                                                                         \m . 
\ee 
\n {\bf Definition 3} The  {\it relational spaces} alias {\it shape-and-scale spaces} are 
\be 
\FrR(N, \mathbb{C}^d)  \:=  \frac{\FrQ(N, \mathbb{C}^d)}{Aut(\langle \FrC, \sigma \rangle)}                                     \m , 
\ee
for $Aut$ denoting automorphisms and $\sigma$ some level of geometrical structure on the carrier space $\FrC^d$.  

\m 
 
\n The relational spaces in the current article are 
\be 
\FrR(N, \mathbb{S}^1)  \:=  \frac{\FrQ(N, \mathbb{S}^1)}{SO(2)} 
                         = \frac{\bigtimes_{i = 1}^N \mathbb{S}^1 }{\mathbb{S}^1} 
						 = \bigtimes_{i = 1}^n \mathbb{S}^1 
						 = \mathbb{T}^n                                     \m : \m \mbox{ the $n$-dimensional torus }          \m ,  
\ee 
for
\be
\mbox{(independent circular relative angle number) } \m n := N - 1                                                              \m .  
\ee 
\n{\bf Remark 4} Quotienting out scale is moreover not possible here, since $\mathbb{S}^1$'s topological identification globally invalidates 
the dilational generator from constituting a generalized (similarity) Killing vector. 
Thus $\mathbb{S}^1$ shape-and-scale theory does not have a meaningful pure (similarity) shape counterpart.

\m 

\n{\bf Remark 5} Relative angles suffice as invariants on $\mathbb{S}^1$; see \cite{AMech, ABook, I, II, III} for further discussion of invariants in Shape(-and-Scale) Theory.  

\m 
 
\n{\bf Remark 6} The {\it minimal relationally nontrivial unit} is the smallest-$N$ configuration with at least 2 degrees of freedom 
(so that one can change with respect to the other). 
For Isometric Shape-and-Scale Theory on $\mathbb{S}^1$, $N = 3$ i.e.\ $n = 2$ is required.  

\m 

\n{\bf Remark 7} To model mirror image and point-or-particle label indistinguishabilities, we furthermore quotient by a discrete group $\Gamma$. 

\m 

\n{\bf Definition 7} If both of these quotientings are carried out to maximal extent -- $\Gamma = S_N \times C_2$ -- one arrives at th {\it Leibniz space}: 
the most discretely quotiented shape(-and-scale) spaces, denoted by $\Leib$.  

\m 

\n{\bf Remark 8} The totality of the possible discrete quotientings form a lattice of distinct discrete subgroup actions (Fig \ref{Lattice}) 
and a corresponding lattice of quotient relational spaces.  
%
{            \begin{figure}[!ht]
\centering
\includegraphics[width=0.32\textwidth]{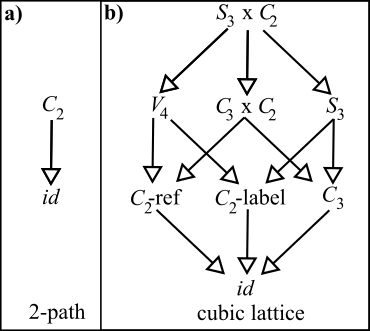}
\caption[Text der im Bilderverzeichnis auftaucht]{        \footnotesize{a) $N = 2$'s and 
                                                                        b) $N = 3$'s lattices of discrete subgroup actions 
																		on the corresponding relational spaces on $\mathbb{S}^1$.
$C_k$ are cyclic groups, $S_N$ are permutation groups, $V_4 = C_2 \times C_2$ and $\times$ denotes the direct product of groups.  
White-headed arrows are reserved in the current article to depict lattices.
} }
\label{Lattice} \end{figure}          }

\vspace{10in}

\section{\biN = 1} 

\n{\bf Remark 1} This case has just the one topologically-distinct configuration O = G (Fig \ref{1-Configs}). 

\m 

\n{\bf Remark 2} By $C(N, N) = 1$, there is one possible labelling for this O and all subsequent further allocations of O.  

\m 

\n{\bf Remark 3} The $N = 1$ O is moreover exceptional in not being a coincidence or a collision, by the model containing no further point to coincide with or particle to collide with. 
%
{            \begin{figure}[!ht]
\centering
\includegraphics[width=0.15\textwidth]{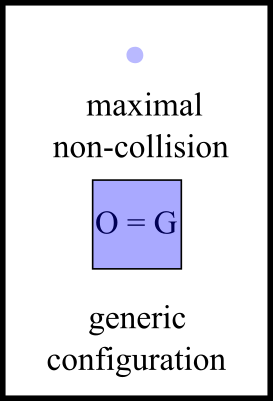}
\caption[Text der im Bilderverzeichnis auftaucht]{        \footnotesize{
\n The sole topological configuration for $N = 1$. } }
\label{1-Configs} \end{figure}          }

\m 
 
\n{\bf Remark 4} For $N = 1$ on arbitrary $\FrC^d$, the constellation space is just 
\be
\FrQ(1, \FrC^d) = \FrC^d                                                                          \m :  
\ee
the carrier space itself. 
Thus in particular for relational theory on the circle, the constellation space is 
\be
\FrQ(1, \mathbb{S}^1) = \mathbb{S}^1                                                              \m . 
\ee 
\n{\bf Remark 5} 
\be 
Isom(\mathbb{S}^1) = SO(2) = U(1) \s{\st}{=} \mathbb{S}^1                                         \m ,
\ee 
where $\s{\st}{=}$ denotes topological equivalence.  

\m 

\n{\bf Remark 6} For $N = 1$ relational theory on the circle, the relational space is 
\be 
\FrR(1, \mathbb{S}^1)  \:=  \frac{\FrQ(1, \mathbb{S}^1)}{SO(2)}
                       \es  \frac{\mathbb{S}^1}{\mathbb{S}^1}
                       \es  id   
                        =   \{  \mbox{pt}  \}  					   \m . 
\ee
This can be envisaged from taking the particle to the circle's `North pole': the axis about which the polar angle $\varphi$ parametrizing the circle is to be measured.  
%
{            \begin{figure}[!ht]
\centering
\includegraphics[width=0.15\textwidth]{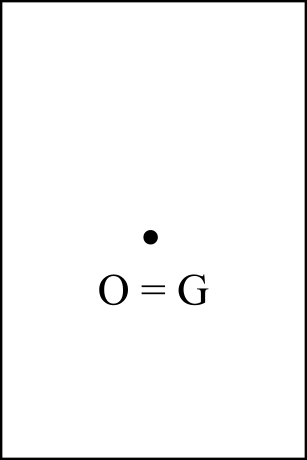}
\caption[Text der im Bilderverzeichnis auftaucht]{        \footnotesize{
\n $N = 1$ relational space; regardless of the model, this is just the same as the particle.} }
\label{1-R} \end{figure}          }

\m 

\n{\bf Remark 7} The only discrete group that can act is $id$, so $N = 1$ supports no further distinct configuration spaces. 

\m 

\n In particular, the $N = 1$ Leibniz relational space for the circle is also just 
\be 
\Leib_{\sFrR}(1, \mathbb{S}^1)  =  \FrR(1, \mathbb{S}^1) 
                                =  \{  \mbox{pt}  \}                        \m.  
\ee

\section{\biN = 2} 

\n{\bf Remark 1} This case has two topologically-distinct configurations \cite{I}: the generic  G and the binary collision B = O (row 1 of Fig \ref{2-Configs}). 
%
{            \begin{figure}[!ht]
\centering
\includegraphics[width=0.45\textwidth]{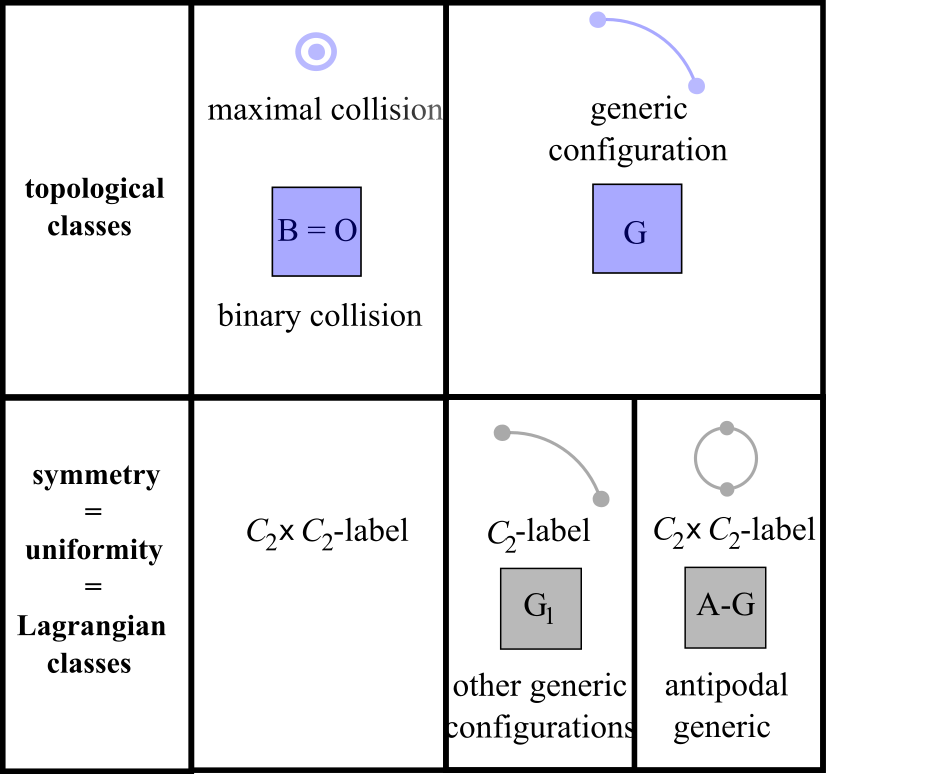}
\caption[Text der im Bilderverzeichnis auftaucht]{        \footnotesize{Row 1 depicts the topologically distinct configurations for $N = 2$, 
with row 2 splitting these up further according to some metric-level features. 
On $\FrC^d =\mathbb{S}^1$, Lagrangian-level distinctions consist of relative angle quantities.} }
\label{2-Configs} \end{figure}          }

\m

\n{\bf Remark 2} The $N = 2$ constellation space on $\mathbb{S}^1$ is 
\be
\FrQ(2, \mathbb{S}^1)  =  \mathbb{S}^1 \times \mathbb{S}^1   
                       =  \mathbb{T}^2                                                                                     \m : \m \mbox{ the 2-torus}  \m .    
\ee
\n{\bf Remark 3} The generic configuration G now comes with two possible labellings: point-or-particle E to the left of F, and F to the left of E.  

\m 

\n{\bf Remark 4} G is parametrized by the relative angle 
\be 
\phi_{12}  :=  \varphi_2 - \varphi_1                                                                                       \m .  
\ee 
\n{\bf Remark 4} The $N = 2$ relational space on $\mathbb{S}^1$ is (Fig \ref{2-RC}.a)
\be 
\FrR(2, \mathbb{S}^1)  \:=  \frac{\FrQ(2, \mathbb{S}^1)}{SO(2)}
                       \es  \frac{\mathbb{S}^1 \times \mathbb{S}^1}{\mathbb{S}^1}
                       \es  \mathbb{S}^1                                                                                   \m . 
\ee
\n{\bf Remark 5} $C_2$-ref and $C_2$-label have the same action on $\FrR(2, \mathbb{S}^1)$, so we just write $C_2$. 					    

\m 

\n{\bf Remark 6} In particular the only other element in the discrete subgroup lattice \ref{Lattice} in this case is  
\be 
\Leib_{\sFrR}(2, \mathbb{S}^1) = \FrR(2, \mathbb{S}^1; C_2)  \:=  \frac{\FrQ(2, \mathbb{S}^1)}{SO(2) \times C_2}
                                       \es  \frac{\mathbb{S}^1 \times \mathbb{S}^1}{\mathbb{S}^1 \times C_2}
                                       \es  \mathbb{S}^1_0                                       
                                       \s{\st}{=} I  									   \m . 
\ee
This is metrically a semicircle $\mathbb{S}^1_0$ and topologically just a closed interval, $I$.  

\m 

\n{\bf Remark 6} The B = O configuration works just as $N = 1$ does (the first of the current article's realizations of \cite{I}'s Relational Aufbau Principle), 
providing a single distinctive point on the circle (Fig \ref{2-RC}.b). 
This is part of the same stratum as the G configuration.  

\m 

\n{\bf Remark 7} The other endpoint of $\Leib_{\sFrR}(2, \mathbb{S}^1)$ -- and boundary between the E-F and F-E label distinctions -- 
is identified at the metric level with being the antipodal configuration A (alias A-G standing for `antipodal generic' once further antipodal configurations enter consideration).
%
{            \begin{figure}[!ht]
\centering
\includegraphics[width=0.35\textwidth]{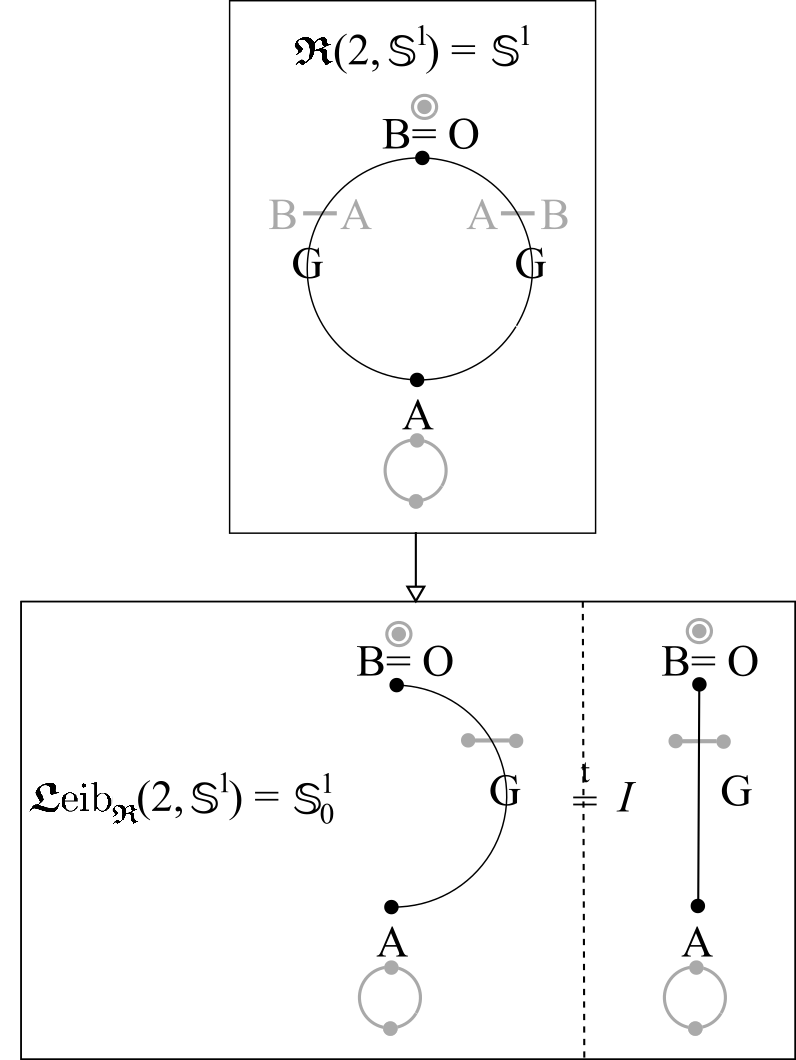}
\caption[Text der im Bilderverzeichnis auftaucht]{        \footnotesize{$N = 2$ relational spaces. 
$\s{\st}{=}$ denotes `equals at the topological level'.} }
\label{2-RC} \end{figure}          }
 
\vspace{10in}
 
\section{\biN = 3} 

\n{\bf Remark 1} There are now three topologically-distinct configurations \cite{III}: 
the generic G, 
the binary collision B, 
and the ternary collision T = O (row 1 of Fig \ref{3-Configs}). 
%
{            \begin{figure}[!ht]
\centering
\includegraphics[width=0.65\textwidth]{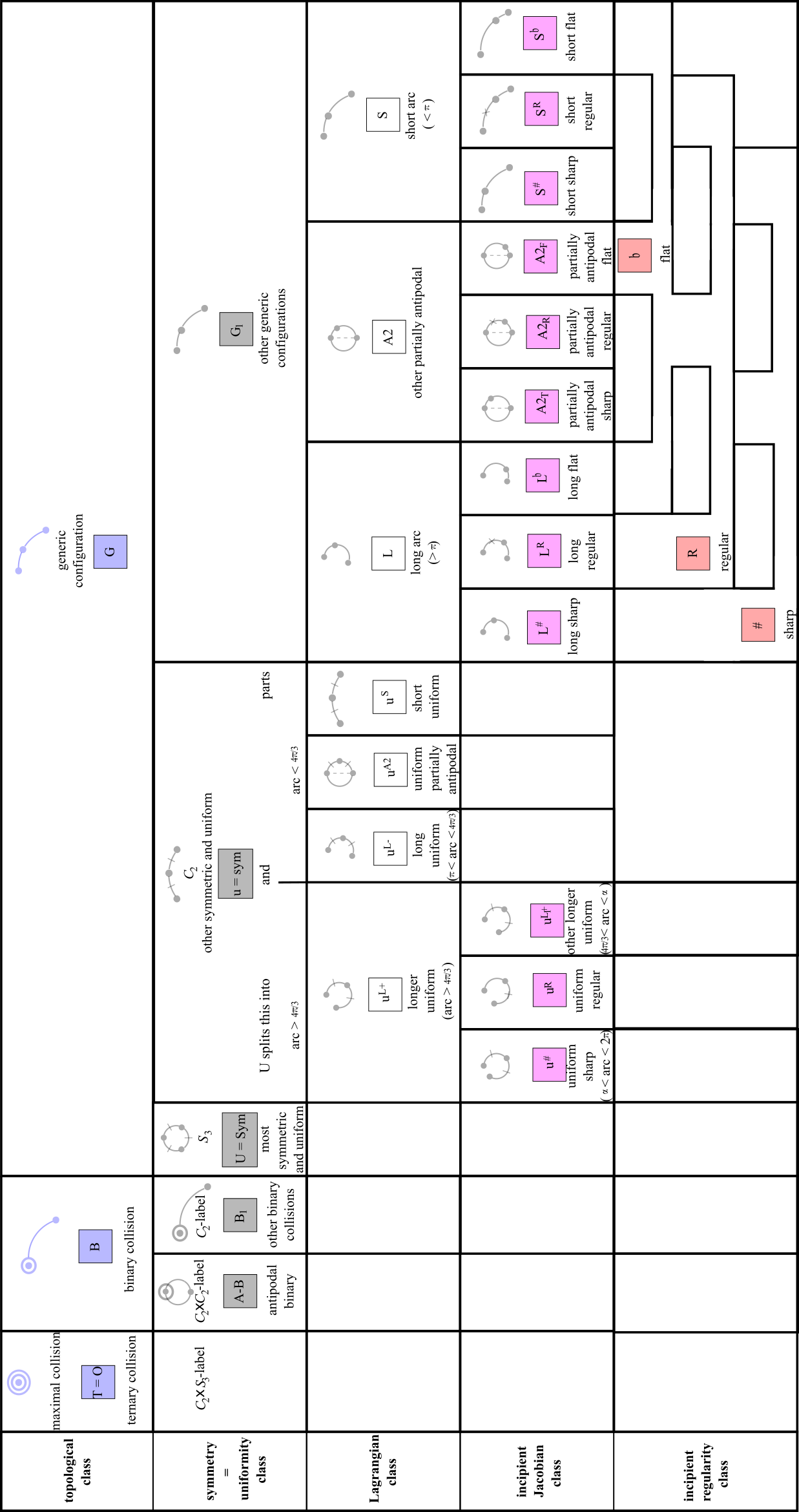}
\caption[Text der im Bilderverzeichnis auftaucht]{        \footnotesize{Row 1 depicts topologically distinct configurations for $N = 3$, 
with row 2 splitting these up further according to symmetry features, row 3 extra Lagrangian features and row 4 for extra incipient Jacobian features.
Row 5 alternatively implements regularity, dividing notions of flatness and sharpness.   
See eq (\ref{a=UR}) for $\alpha$'s value and meaning. 
} }
\label{3-Configs} \end{figure}          }

\m

\n{\bf Remark 2} The $N = 3$ constellation space on the circle is 
\be
\FrQ(3, \mathbb{S}^1)  \es  \bigtimes_{i = 1}^3 \mathbb{S}^1 = \mathbb{T}^3       \m : \mbox{ the 3-torus}                          \m .  
\ee
\n{\bf Remark 3} The $N = 3$ relational space on the circle is 
\be 
\FrR(3, \mathbb{S}^1)  \:=  \frac{  \FrQ(3, \mathbb{S}^1)  }{  SO(2)  }
                       \es  \frac{  \bigtimes_{i = 1}^3 \mathbb{S}^1  }{  \mathbb{S}^1  }
                       \es  \mathbb{S}^1 \times \mathbb{S}^1
                        =	\mathbb{T}^2				                          \m : \mbox{ the 2-torus}                           \m . 
\ee

\n{\bf Remark 4} Let us examine this further at the metric level.  
Firstly $\FrR(3, 1)$ is as depicted in Fig \ref{R(3,1)-Metric-Lattice}.1). 
In fact, 
\be 
\FrR(3, 1; C_3) \:= \frac{\FrR(3, 1)}{C_3} 
\ee 
as depicted in Fig \ref{R(3,1)-Metric-Lattice}.3) is one step closer to what we need, by cycling points-or-particles around the circle.  

\m 

\n{\bf Remark 5} Secondly, we topologically identify to form the representation of $\FrR(3, \mathbb{S}^1)$ of Fig \ref{Torus-Prelim}.c). 
Note that metrically this is not a square or even a rectangle, but rather a parallelogram; it remains however a valid presentation of the torus.
This rests on passage from Fig \ref{Torus-Prelim}.a)'s absolute angles relative to a fixed axis NS (`North--South') 
to the relational Jacobi angle variables of Fig \ref{Torus-Prelim}.b).
These are, algebraically, 
\be 
\theta_1 = \varphi_2 - \varphi_1                           \m , 
\ee
\be 
\theta_2 = \varphi_3 - \frac{1}{2}(\varphi_1 + \varphi_2)  \m , 
\ee  
in the case of equal particle masses assumed in the current article. 
Note that these are directly analogous to relative Jacobi coordinates except that they are now periodic variables, with ranges 0 to $2 \, \pi$. 
We next Jacobi mass-weight these to obtain 
\be 
\slTheta_1 = \frac{1}{\sqrt{2}} \, \theta_1   \m , 
\ee 
\be 
\slTheta_2 = \sqrt{\frac{2}{3}} \, \theta_2  \m .  
\ee 
A virtue of these coordinates is that they are flat Cartesian coordinates. 
With their ranges being slightly cumbersome, however, we proportionately rescale these variables using the scalefactor 
\be 
k = \frac{1}{\pi} \sqrt{\frac{3}{2}}
\ee 
to obtain our final variables
\be 
\Theta_1 \m \mbox{ with range 0 to $\sqrt{3}$} \m \mbox{ and}
\ee
\be 
\Theta_2 \m \mbox{ with range 0 to 2}          \m .  
\ee  
\n{\bf Remark 6} While one is most familiar with square or rectangular identification tori having 2 commuting Killing vectors perpendicular to the two sides, 
this continues to occur for parallelogram-identified tori. 
Thus 
\be 
Isom(\FrR(3, \mathbb{S}^1))     =       Isom(\mathbb{T}^2)  
                               =       U(1) \times U(1)
                          \m [ \m \s{\st}{=}  \mathbb{S}^1 \times \mathbb{S}^1 
                               =       \mathbb{T}^2        					\m ] \m	   
\ee 
is valid for tori made with this identification as well.
In terms of $\Theta_1$ and $\Theta_2$, generator Killing vectors for this are
\be 
\frac{\pa}{\pa\Theta_1} \m \mbox{ and } \m   \frac{\pa}{\pa \Theta_1} + \sqrt{3} \frac{\pa}{\pa \Theta_2} \m . 
\ee 
These are globally valid across the topologically identification, periodic, and clearly commute with each other. 

\m 

\n{\bf Remark 7} Within the fundamental cell parallelogram described, the B configurations correspond to (Fig 7.c) 
\be
\Theta_1 = 0         \m ,
\ee
\be
\Theta_1 = \sqrt{3} \, \Theta_2         \mma \mbox{and}
\ee 
\be 
\Theta_1 = - \sqrt{3} \, \Theta_2  \m .
\ee
These split the parallelogram (or subsequently identified torus) into a 2-tile tesellation as indicated in Fig 7.c) in green and cyan.
%
{            \begin{figure}[!ht]
\centering
\includegraphics[width=0.6\textwidth]{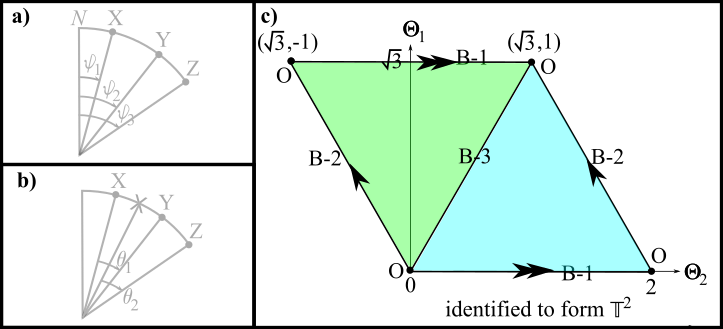}
\caption[Text der im Bilderverzeichnis auftaucht]{        \footnotesize{a) Absolute angles. b) Relative Jacobi angles; note that there is no longer any reference to $N$. 
c) The identified parallelogram model of $\FrR(3, \mathbb{S}^1) = \mathbb{T}^2$ in rescaled mass-weighted relative Jacobi angle coordinates $\Theta_1, \Theta_2$.  } }
\label{Torus-Prelim} \end{figure}          }

\n{\bf Structure 1} All the edges and corners of the fundamental cell parallelogram are known. 
To deal with discrete quotients, however, we additionally need the symmetry structure.

\m 

\n The A, U and u configurations of row 2 of Fig 6 are relevant to this. 
We count out that there are 3 A points, two U points and six u lines as per Fig \ref{C-Anti-Sym-Lag}.1)'s 12-tile tessellation.  
%
{            \begin{figure}[!ht]
\centering
\includegraphics[width=1.0\textwidth]{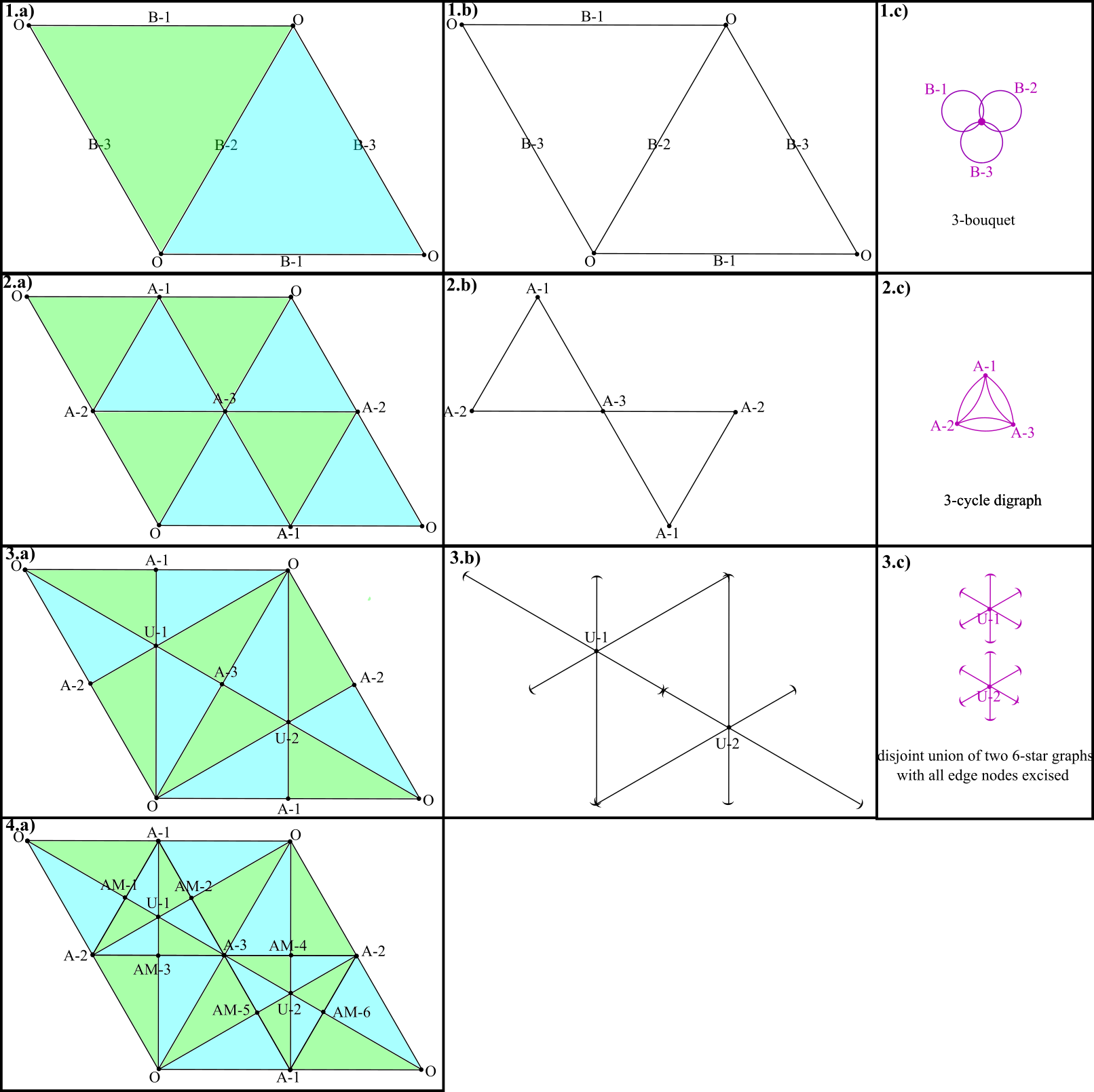}
\caption[Text der im Bilderverzeichnis auftaucht]{        \footnotesize{
\n From 1.a), the collision sructure of 1.b) can be read off; the corresponding topological-level graph is the 3-bouquet of Fig 1.c).  

\m

\n 2.a) The symmetry structure, here aligned entirely with uniformity structure gives a 12-tile  tessellation of the shape-and-scale space torus. 
The uniformity structure itself is as in 2.b) (coincidences-or-collisions are excised from our definition of uniformity \cite{I}; 
this is the partly deleted-vertex graph of 2.c). 

\m 

\n 3.a) Antipodal configurations give an 8-tile tessellation of the shape-and-scale space torus. 
The antipodal structure itself is as in 3.b); this is not yet a graph because each A point appears twice; taking this into account gives the digraph of Fig 3.c). 

\m 

\n 4.a) Combining these structures, the Lagrangian structure forms the indicated 24-tile tessellation of the shape-and-scale space torus. 
In this case, the faces are part of the structure, so there is no need for auxiliary subfigures 4.b) or 4.c). 
AM is a shorter notation for Fig \ref{3-Configs}'s $A2_{\sR}$, which anticipates the Jacobian-level meaning: M for merger.)
} }
\label{C-Anti-Sym-Lag} \end{figure}          }

\m 

\n This suffices to know all the edges that occur in forming the discrete quotient lattice of Fig \ref{3C-Latt}.  

{            \begin{figure}[!ht]
\centering
\includegraphics[width=0.7\textwidth]{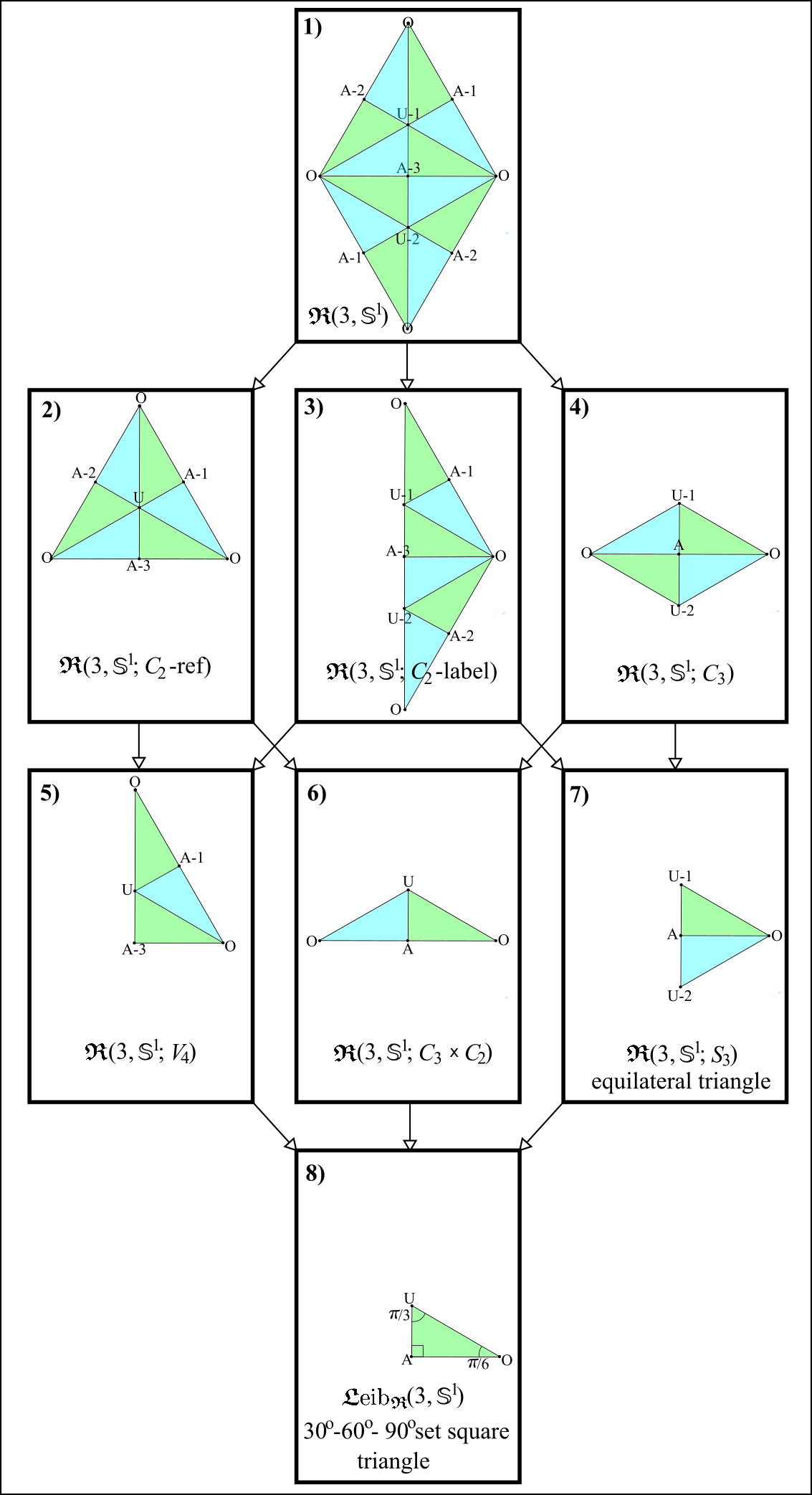}
\caption[Text der im Bilderverzeichnis auftaucht]{        \footnotesize{Lattice of discrete quotient shape-and-scale spaces.  
} }
\label{3C-Latt} \end{figure}          }

\m

\n {\bf Remark 8} Quotienting out by $C_2$-ref works out as per Fig \ref{3C-Latt}.2) and \ref{3-R}.2), folding along its A2 lines depicted in the latter to give a tetrahaedron, 
i.e.\ topologically a sphere $\mathbb{S}^2$.

\m 

\n {\bf Remark 9} Quotienting out by $C_2$-label works out as per Fig \ref{3C-Latt}.3) and \ref{3-R}.3), 
folding about its horizontal B line to give a cone, followed by back-to-front identification of its edge circle to give $\mathbb{RP}^2$ topologically.

\m 

\n {\bf Remark 10} Quotienting out by $C_3$ works out as per Fig \ref{3C-Latt}.4) and \ref{3-R}.4), 
folding about the U-1 to U-2 line to give a cone with angle $2 \, \pi/3$ about the point O. 
This is topologically a sphere $\mathbb{S}^2$. 
%
{            \begin{figure}[!ht]
\centering
\includegraphics[width=0.7\textwidth]{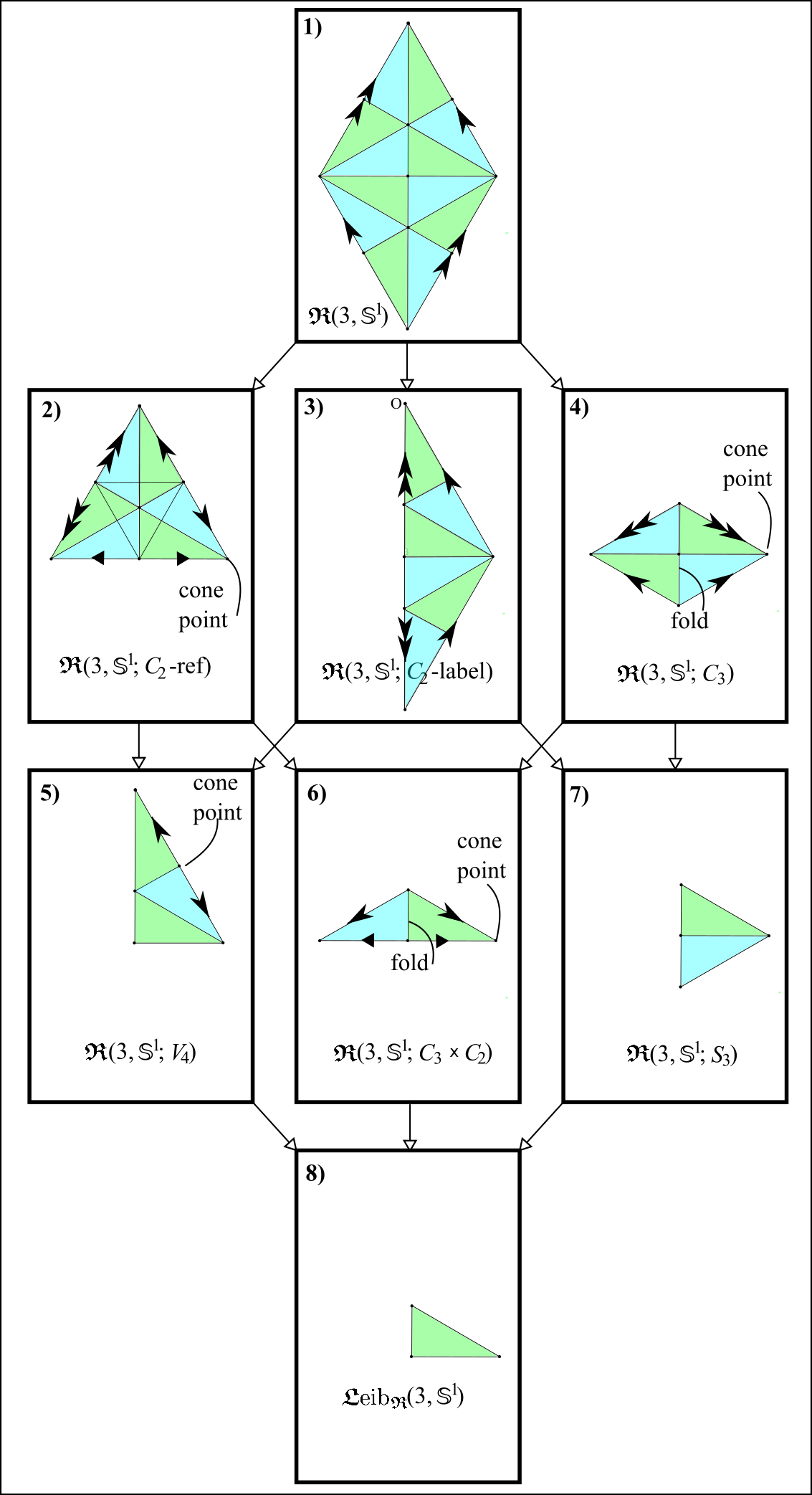}
\caption[Text der im Bilderverzeichnis auftaucht]{        \footnotesize{Topological identifications for Relational spaces for 3 points-or-particles on the circle. } }
\label{3-R} \end{figure}          }

\m 

\n {\bf Remark 11} Quotienting out by $V_4 = C_2 \times C_2$ works out as per Fig \ref{3C-Latt}.5) and \ref{3-R}.5), 
giving a cone with edge circle, which is topologically a disc $\mathbb{D}^2$.

\m 

\n {\bf Remark 12} Quotienting out by $C_3 \times C_2$ works out as per Fig \ref{3C-Latt}.6) and \ref{3-R}.6), 
folding about the A to U line to give a cone with angle $\pi/3$ about the point O. 
This is once again topologically a sphere $\mathbb{S}^2$.  

\m

\n {\bf Remark 13} Quotienting out by $S_3$ works out as per Fig \ref{3C-Latt}.7) or \ref{3-R}.7), 
giving an equilateral triangle region (topologically a disc $\mathbb{D}^2$).   

\m

\n {\bf Remark 14} Quotienting out by $S_3 \times C_2$ gives the Leibniz space as per Fig \ref{3C-Latt}.8) or \ref{3-R}.8), 
giving a $30^{\so}$-$60^{\so}$-90$^{\so}$ set square' triangle region (topologically a disc $\mathbb{D}^2$).   

\m 

\n{\bf Remark 15} We summarize the final results in lattice form in Fig \ref{Top-Latt}.   
%
{            \begin{figure}[!ht]
\centering
\includegraphics[width=0.4\textwidth]{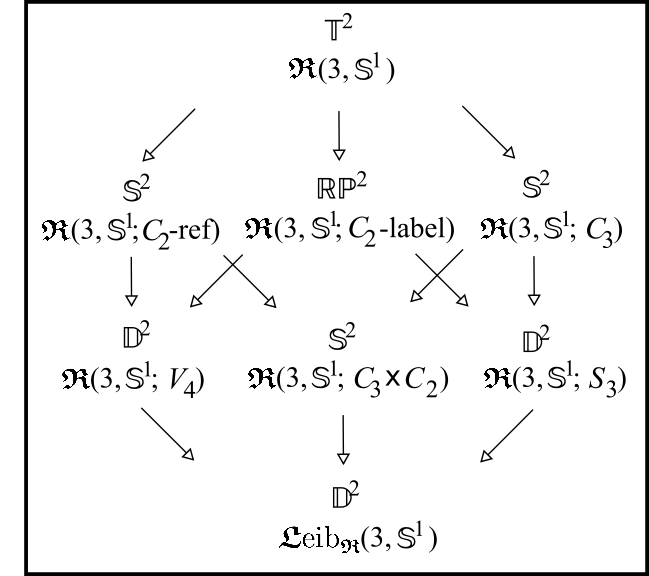}
\caption[Text der im Bilderverzeichnis auftaucht]{        \footnotesize{Relational spaces for 3 points-or-particles on the circle at the topological level. } }
\label{Top-Latt} \end{figure}          }

\section{Further structures}

\n{\bf Structure 1} $\FrR(3, \mathbb{S}^1)$ moreover also possesses an {\it antipodal structure} consisting of totally antipodal A points and partially antipodal A2 lines 
as depicted in Fig \ref{C-Anti-Sym-Lag}.2). 
This marks the parallelogram/torus with an 8-tile tessellation. 

\m 

\n{\bf Structure 2} Combining the previous two structures in the current case suffices to obtain the {\it Lagrangian structure}: 
picking out all configurations which can be expressed in terms of zero or equal relative angles.  
This forms Fig \ref{C-Anti-Sym-Lag}.3)'s 24-tile tessellation.   

\m

\n{\bf Remark 1} Having obtained $\Leib_{\sFrR}(3, \mathbb{S}^1)$, 
we give its topological, coincidence-or-collision, symmetry and uniformity structure in Figs \ref{Leib(3,S1)-Top}) and 13.2.b), noting morevoer that the last two of these 
are the same in this example.    
Topology-and-symmetry structure is included as well, out of its sufficing to define all of $\Leib$'s perimeter.

\m 

\n{\bf Remark 2}  The full Lagrangian structure's edges form the triangulation of Fig \ref{Leib(3,S1)-Lag}.a), whose edges form the gem graph, 
                                                        and the ajacency graph of Fig \ref{Leib(3,S1)-Lag}.b).  

\m 

\n{\bf Remark 7} The full Lagangian structure of the unquotiented version is in Fig \ref{C-Anti-Sym-Lag}.4.a).

\m 

\n{\bf Remark 8} Within this, an 8-tessellation by antipodal structure and a 12-tessellation by topological-and-symmetric structure can be observed.  

\m 

\n{\bf Remark 9} The O configuration works just as $N = 1$ does, providing a single distinctive point on the circle (Fig \ref{3-R}). 

\m 

\n{\bf Remark 10} We have 1 O and 3 B's (like for 3 particles in $\mathbb{R}^2$ \cite{M05, +Tri}).  

\m 

\n{\bf Remark 11} The A-B configuration works just as $N = 2$'s antipodal configuration does, providing a single distinctive point on the circle (Fig \ref{3-R}). 
\n There are moreover three labellings of B and of A-B to one of A.

\vspace{10in}

{            \begin{figure}[!ht]
\centering
\includegraphics[width=0.8\textwidth]{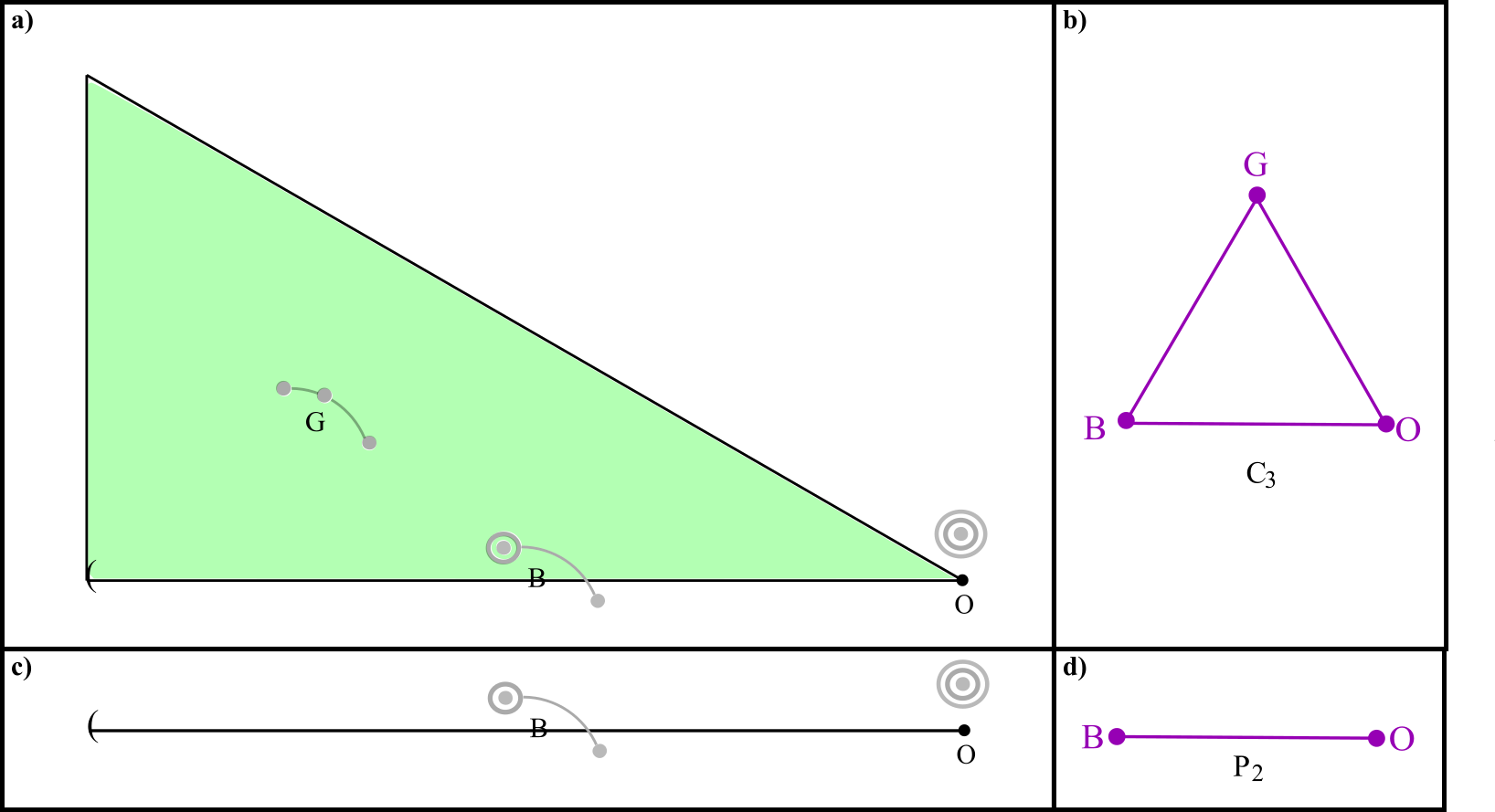}
\caption[Text der im Bilderverzeichnis auftaucht]{        \footnotesize{a) $(\Top(\Leib_{\tFrR}(3, \mathbb{S}^1))$: 
the Leibniz space for 3 points-or-particles in $\mathbb{S}^1$ at the topological level.

\m

\n b) The corresponding adjacency graph, $\FrA((\Top \bigcup \Top)(\Leib_{\tFrS}(3, \mathbb{S}^1)))$; 
at the level of unlabelled graphs, this is the 3-cycle graph $\mC_3$.

\m 

\n c) $(\Co(\Leib_{\tFrR}(3, \mathbb{S}^1))$: the Leibniz space for 3 points-or-particles in $\mathbb{S}^1$'s coincidence-or-collision structure.

\m

\n d) The corresponding adjacency graph, $\FrA((\Co \bigcup \Top)(\Leib_{\tFrS}(3, \mathbb{S}^1)))$; at the level of unlabelled graphs, this is just the 2-path graph $\mP_2$.
} }
\label{Leib(3,S1)-Top} \end{figure}          }
%
{            \begin{figure}[!ht]
\centering
\includegraphics[width=1\textwidth]{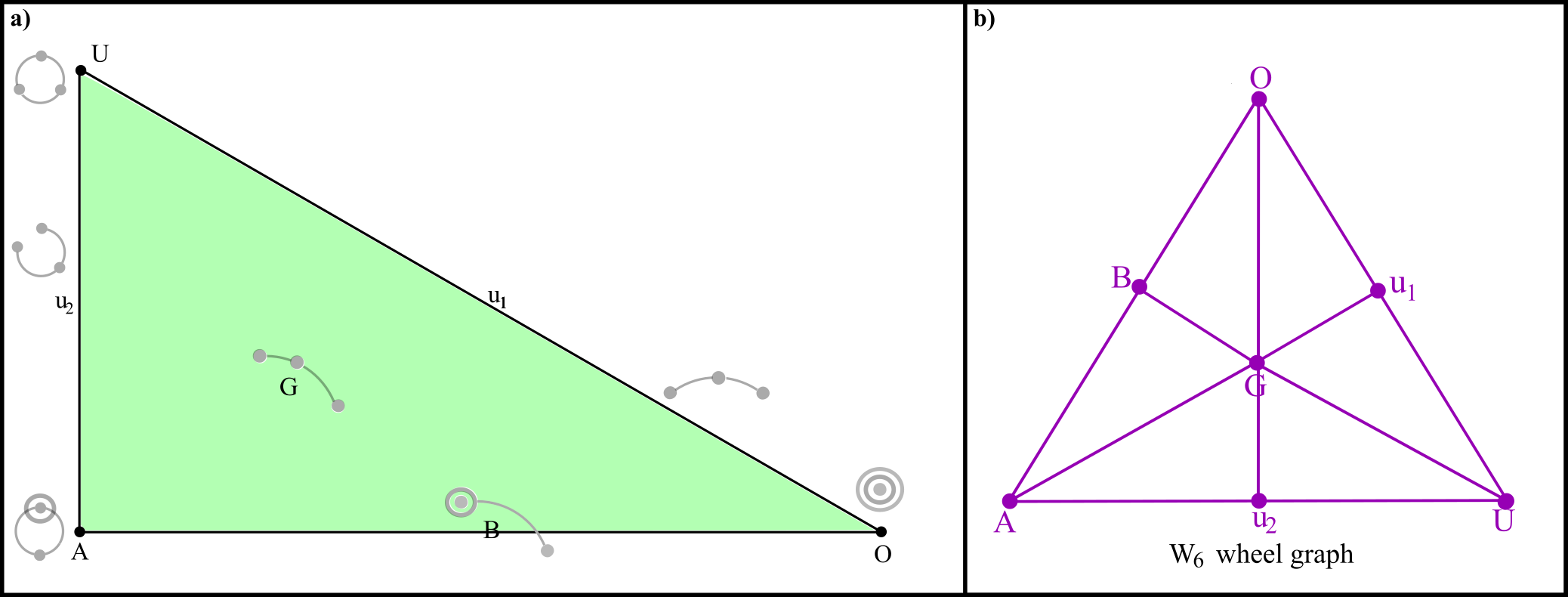}
\caption[Text der im Bilderverzeichnis auftaucht]{        \footnotesize{a) $(\Sym \bigcup \Top)(\Leib_{\tFrR}(3, \mathbb{S}^1)) 
                                                                          = (\Uni \bigcup \Top)(\Leib_{\tFrS}(3, \mathbb{S}^1)))$: 
the Leibniz space for 3 points-or-particles in $\mathbb{S}^1$ at the level of symmetry as well as topology, or, alternatively, of (Lagrangian) uniformity instead of symmetry.   

\m

\n b) The corresponding adjacency graph, $\FrA((\Sym \bigcup \Top)(\Leib_{\tFrS}(3, \mathbb{S}^1)))$; 
at the level of unlabelled graphs, this is the 6-spoked wheel graph $W_6$.} }
\label{Leib(3,S1)-Sym} \end{figure}          }

\vspace{10in}
 
{            \begin{figure}[!ht]
\centering
\includegraphics[width=1\textwidth]{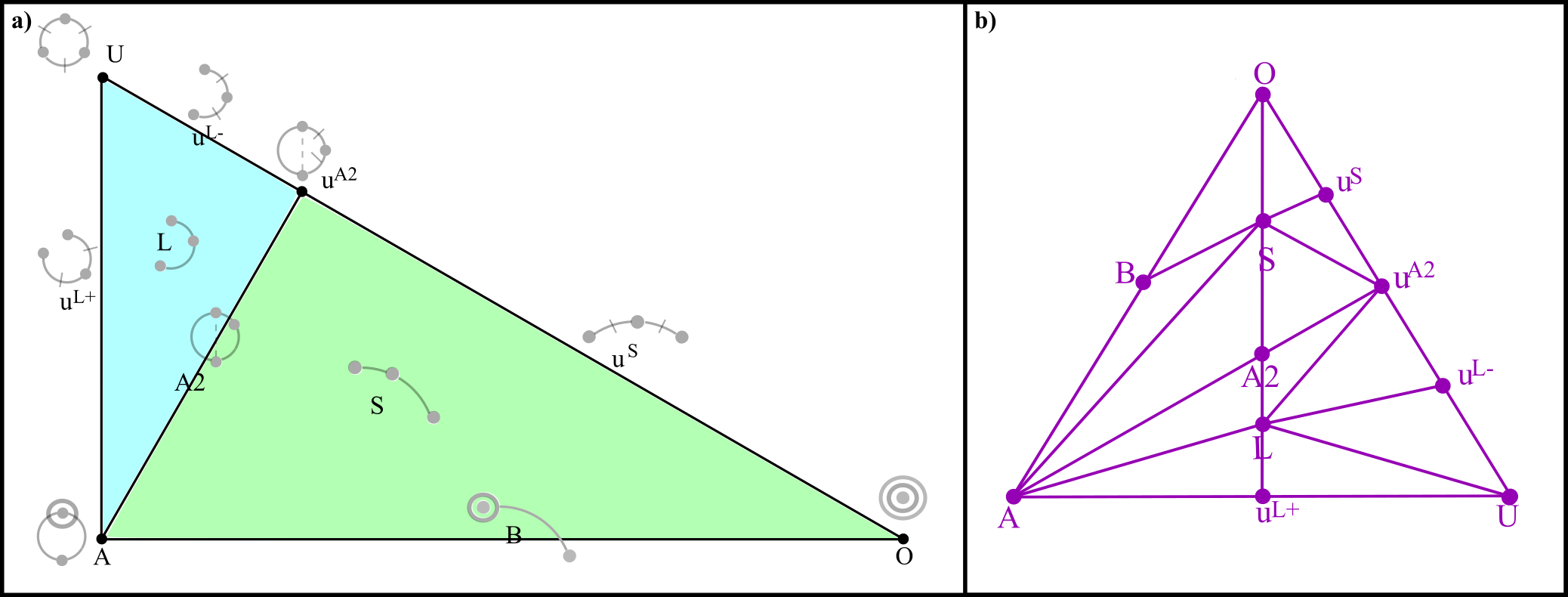}
\caption[Text der im Bilderverzeichnis auftaucht]{        \footnotesize{a) $\Lag(\Leib_{\tFrR}(3, \mathbb{S}^1))$: 
the Leibniz space for 3 points-or-particles in $\mathbb{S}^1$ at the Lagrangian level, which amounts to further including the rightness structure.  

\m 

\n b) The corresponding adjacency graph, $\FrA(\Lag(\Leib_{\tFrR}(3, \mathbb{S}^1)))$.   
At the level of unlabelled graphs, this is two 6-spoked wheel graphs $W_6$ joined along two adjacent-perimeter edges.} }
\label{Leib(3,S1)-Lag} \end{figure}          }
%
{            \begin{figure}[!ht]
\centering
\includegraphics[width=1.0\textwidth]{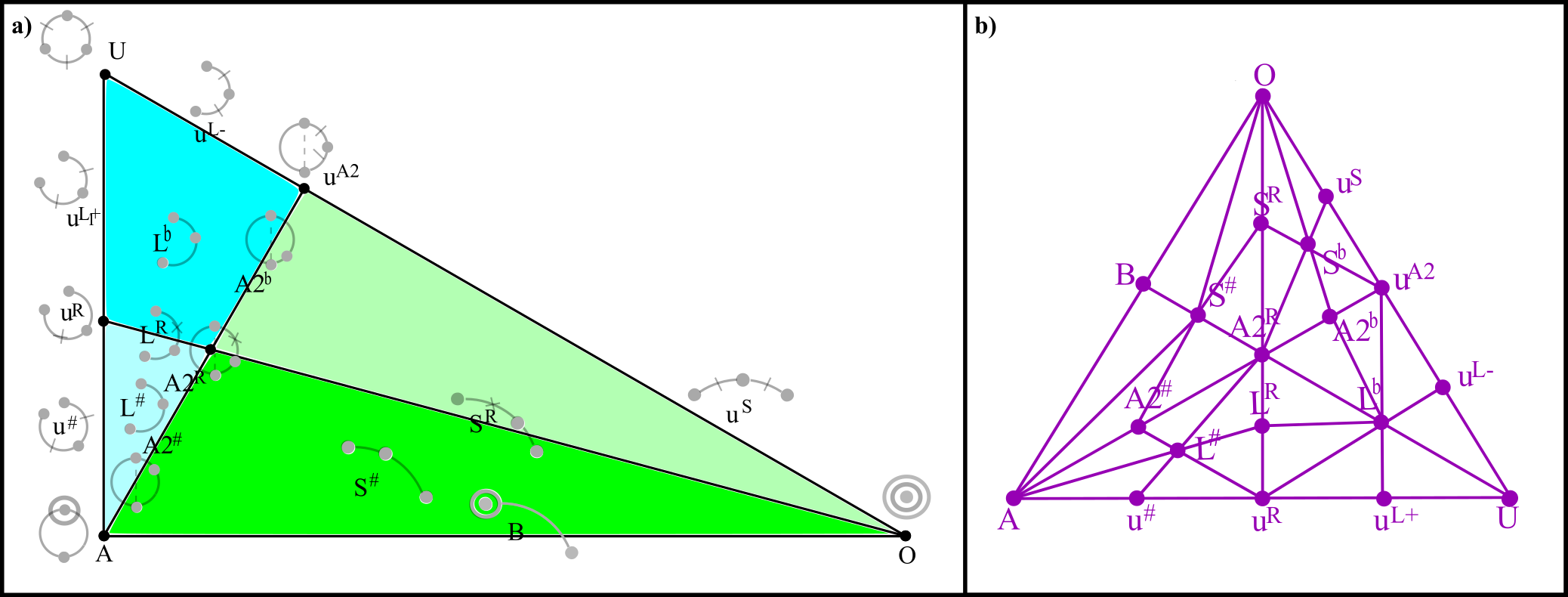}
\caption[Text der im Bilderverzeichnis auftaucht]{        \footnotesize{a) $\Incip\mbox{-}\Jac(\Leib_{\tFrR}(3, \mathbb{S}^1))$:  
the Leibniz space for 3 points-or-particles in $\mathbb{S}^1$ at the Jacobian level structure, which amounts to further including the regularity structure.  

\m 

\n b)  The corresponding adjacency graph $\FrA(\Incip\mbox{-}\Jac(\Leib_{\tFrR}(3, \mathbb{S}^1)))$.
At the level of unlabelled graphs, this is three 6-spoked wheel graphs $\mW_6$ and one 8-spoked wheel graph $\mW_8$, joined in a cycle as shown 
with two adjacent-perimeter edge joins, all four of which share a common vertex: the regular partially antipodal configuration.}}
\label{Leib(3,S1)-Jac} \end{figure}          }

\vspace{10in}

{            \begin{figure}[!ht]
\centering
\includegraphics[width=1.0\textwidth]{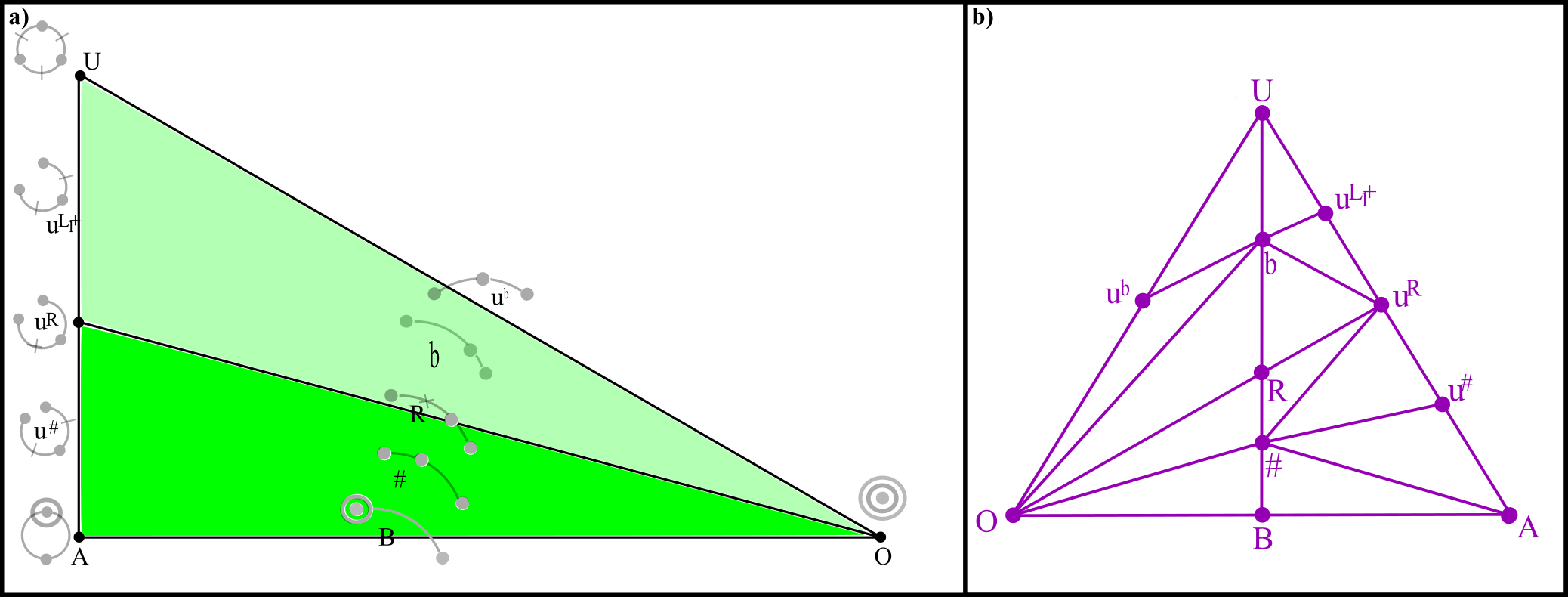}
\caption[Text der im Bilderverzeichnis auftaucht]{        \footnotesize{a) $\Reg(\Leib_{\tFrS}(3, 2))$: 
the Leibniz space for 3 points-or-particles in $\mathbb{S}^1$ including the regularity structure but not the rightness structure.

\m

\n b) The corresponding adjacency graph $\FrA(\Reg(\Leib_{\tFrR}(3, 2)))$; this is clearly isomorphic at the level of unlabelled graphs with Fig \ref{Leib(3,S1)-Lag}.b)'s.}}
\label{Leib(3,S1)-Reg} \end{figure}          }

\m 

\n{\bf Remark 12} The uniform-and-regular configuration occurs at arc length
\be 
\alpha := 2 \, \pi \left( 1 -\frac{\sqrt{2}}{5} \right) \approx 1.434 \, \pi \approx 4.506 \m . 
\label{a=UR}
\ee 

\m 
    
\n{\bf Remark 13} Appendices A and B provide comparison with the (3, 1) and (3, 2) models respectively.

\section{Dynamics application} 

\n{\bf Remark 1} Free motions are geodesics on configuration space. 

\m 

\n{\bf Remark 2} For spaces with flat metric, these are straight lines.    

\m 

\n{\bf Remark 3} In the case of the flat-space torus, these line continues straight through on the other side of each topological identification they hit. 
There is a well-known distinction between periodic solutions and solutions which sweep out the entire torus \cite{Arnold}. 
In the square model, these admit a rational versus irrational slope characterization, which can readily be mapped to characterizations for the alternative 
rectangle and parallelogram identification models of the torus.

\m 

\n{\bf Remark 4} The Leibniz space itself is a 2-$d$ closed region. 

\m 

\n{\bf Remark 5} If treated in isolation, reflection boundary conditions apply at edges.  

\m 

\n 0) A geodesic striking an edge at right angles bounces back along the same trajectory. 

\m 

\n 1) A geodesic striking an edge at any other angle is reflected along a distinct geodesic in accordance with 
\be
\mbox{(angle of incidence)} = \mbox{(angle of reflection)} \m ,  
\label{IR}
\ee 
each defined relative to the normal at the point of contact.  
[0) is then just a limiting subcase of this.]

\m 

\n{\bf Remark 6} Specifying boundary conditions for geodesics striking corners turns out to be more complicated.  

\m 

\n 2) A geodesic striking a corner along the line of bisection bounces back along the same trajectory. 

\m 

\n 3) A geodesic striking a corner at any other angle is harder to handle in the present context. 
This is simplest to envision by considering the relational space tessellation by Leibniz spaces. 
It is now possible for 

\m 

\n i) such a trajectory to be reflected according to (\ref{IR}) now defined relative to the angle bisector at the corner -- {\it reflective prolongation} -- or 

\m 

\n ii) sent back along the original trajectory. 

\m 

\n This is illustrated by examples of tessellations in Fig \ref{Dyn-Comp}.  
[2) is then a limiting subcase for which these two boundary conditions cease to be distinct.]

\m 

\n{\bf Remark 6} To obtain an entire trajectory, one follows it in both directions until each has undergone a collision of type 0), 2) or 3.ii).  
The alternative interpretation that types 0), 2) or 3.ii) are absorptive can also be entertained for some purposes 
(intrinsic to the Leibniz space rather than in considering it as the individual tile within a larger tessellation).  

\m 

\n{\bf Remark 7} The partial antipodal A2 line realizes a right-angled collision with a boundary: type 2. 
Its other end collides with a corner realizing type 3.ii)'s reflective prolongation.
By this, Fig \ref{3C-Latt}.1)'s picture of the Lagrangian structure is complete.

\m 

\n{\bf Remark 8} The regularity line emanates from the O corner as a line of bisection -- type 2 -- but strikes the `long-arc uniformity' opposite edge at an oblique angle. 
This is why we term (\ref{Leib(3,S1)-Jac}) {\sl incipient} Jacobi structure. 

\m 

\n{\bf Remark 9} The current article's model is more complicated in this respect than either (3, 2) \cite{III} or (4, 1) \cite{II}, indeed necessitating the above {\sl upgrade} 
on discussions of boundary conditions in these other papers.  
For (3, 2), the lines of rightness and regularity {\sl both} strike their opposite edges at right angles (Figs \ref{Leib(3,S1)-Lag} and \ref{Leib(3,S1)-Reg}), 
by which these are self-contained models of right and regular triangles.
For (4, 1), one line of Jacobi mergers strikes its opposite boundary at right angles (Fig \ref{Dyn-Comp}.c), 
whereas the other undergoes two reflections in the manner of the swallowtail (Fig \ref{Dyn-Comp}.d) before striking a distinct corner in the manner of type 3.i). 
%
{            \begin{figure}[!ht]
\centering
\includegraphics[width=0.9\textwidth]{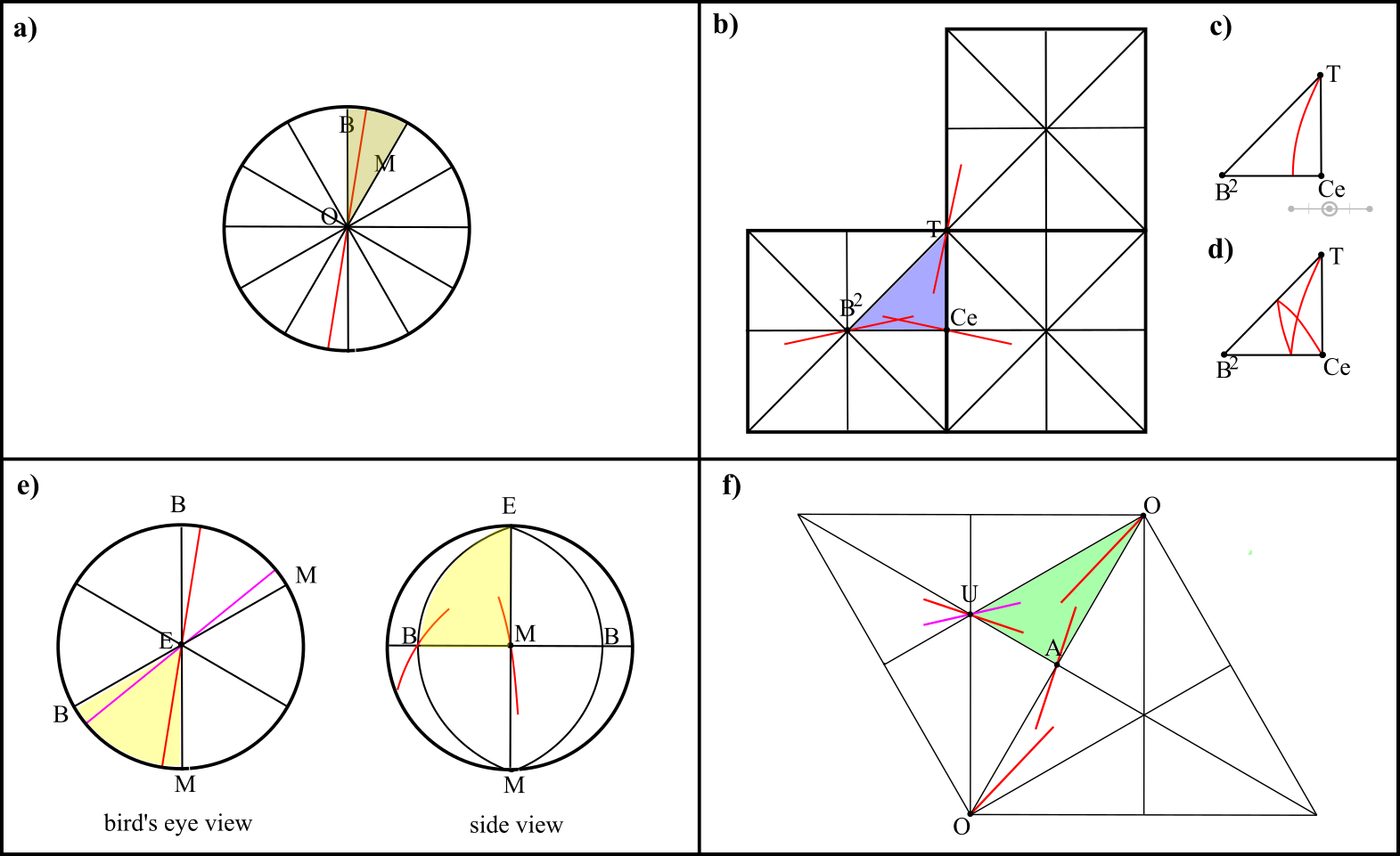}
\caption[Text der im Bilderverzeichnis auftaucht]{        \footnotesize{Boundary conditions for Leibniz spaces (themselves shaded in brown, blue, yellow and green  
for the following four models respectively). 

\m 

\n a) For $\FrR(3, 1)$, the only vertex O is absorptive, 
b) as are all three vertices for $\FrS(4, 1)$ for which c) and d) provide the Jacobi structure.  
The Ce configuration -- central binary -- is depicted in grey at the bottom of c).  

\m 

\n e) For $\FrS(3, 2)$, B and M vertices are absorptive but the equilateral vertex E is nontrivially reflective. 
However, the great circle arc of regularity R is exceptionally absorptive since it bisects the angle at E, 
by which Fig \ref{Leib(3, 2)-Full-Jac} is the totality of the Jacobi structure in this case. 

\m 

\n f) For $\FrS(3, \mathbb{S}^1)$, the O and A vertices are absorptive but there is reflective prolongation at the U vertex 
(the trajectory's return is indicated in magenta rather than red).   
While this is in 1 : 1 correspondence with $\FrS(3, 2)$, the incipient line of regularity for $\FrS(3, \mathbb{S}^1)$ no longer meets the opposite side at right angles. 
This gives a metric-level reason for 3 points-or-particles on a circle being more complicated at the reduced configuration space level than the study of triangles.  
} }
\label{Dyn-Comp} \end{figure}          }

\m

\m 

\m

\m 

\m  

\m  

\m

\m  

\m  

\m 

\m

\m 

\m  

\m  

\m 

\m

\m 

\m  

\m  

\m 

\m

\m 

\m  

\m  

\m 

\m

\m 

\m  

\m  

\m 

\m

\m 

\m  

\m  

\m 

\m

\m 

\m  

\m  

\m 

\m

\m 

\m  

\m  

\m 

\m

\m 

\m  

\m  

\m

\vspace{10in}

\section{Probability application} 

3 points-or-particles on $\mathbb{S}^1$ affords an analogue of shape-theoretic calculation \cite{Small, MIT, A-Pillow} 
of Lewis Carroll's Prob(obtuse) \cite{Carroll, Guy-Portnoy}, for a triangle.
It is moreover simpler to calculate for $\mathbb{S}^1$: a flat triangle Leibniz space rather than a spherical triangle one.  
This gives 
\be
\mbox{Prob}( \, \mbox{arc } < \pi)  \es  \frac{3}{4}                                                                    \m , 
\ee 
as follows from $\Leib_{\sFrR}(3, \mathbb{S}^1)$ possessing the flat metric and corresponding area measure, 
and the partial antipodal line A2 forming similar triangles in a 3 : 1 area ratio.  

\m 

\n Similarly, subdividing up the u part of the perimeter -- of total length 3 -- gives
\be 
\mbox{Prob}\left( \, \mbox{arc } < \pi \, | \, \mbox{uniform} \right)  \es  \frac{\frac{3}{2}}{3} 
                                                                   \es  \frac{1}{2}                                 \m ,
\ee 
\be 
\mbox{Prob}\left( \, \pi < \mbox{arc } < \frac{4 \, \pi}{3} \, \mbox{\Large |} \, \mbox{uniform} \right)  \es  \frac{\frac{1}{2}}{3} 
                                                                                        \es  \frac{1}{6}            \m , 
\ee 
and 
\be 
\mbox{Prob}\left(\mbox{arc } > \frac{4 \, \pi}{3} \, \mbox{\Large |} \, \mbox{uniform} \right)   \es  \frac{1}{3}                  \m .
\ee  
\cite{A-Pillow}'s extra cases of geometrical probabilities -- involving regularity and its combination with rightness and with isoscelesness -- 
also admits a 3 points-or-particles on $\mathbb{S}^1$ analogue.  
Firstly, 
\be 
\eta := \mbox{Prob}(\sharp)    \es    \frac{\frac{3}{2} \tan \, \frac{\pi}{12}}{\frac{\sqrt{3}}{2}} 
                       \es    \sqrt{3} \tan \, \frac{\pi}{12} 
					 \approx 0.464                                                                        \m , 
\ee
and so 
\be 
\mbox{Prob}(\flat)     =     1 - \mbox{Prob}(\sharp) 
                       =     1 - \eta    
				    \approx 0.536                                                                         \m . 
\ee
We made use here of  
\be 
\mbox{Area}(\Leib_{\sFrR}(3, \mathbb{S}^1)) = \frac{\sqrt{3}}{2}                                                 \m .
\ee 
Secondly, using the $\mO \, \mA2^{\sR} \, \muu^{\sA 2}$ right-angled triangle, 
\be 
\mbox{Area}(\mS^{\flat})  \es  \frac{1}{2} \times \frac{3}{2} \times{3}{2} \, \tan \, \frac{\pi}{12}   
                          \es  \frac{9}{8} \, \tan \, \frac{\pi}{12}         \m ,
\ee 
so 
\be 
\mbox{Prob}(\mS^{\flat}) \es  \frac{  \frac{9}{8} \, \tan \, \frac{\pi}{12}  }{\frac{\sqrt{3}}{2}} 
                         \es  \frac{3}{4} \, \eta  
						 \es 3 \, \chi
						\approx 0.348                                                                     \m .
\ee 
$\chi$ is here the most convenient parameter to use to characterize the current suite of probabilities, defined by 
\be 
\chi  :=  \frac{\delta}{4} 
      \es \frac{\sqrt{3}}{4} \, \tan \, \frac{\pi}{12}  
	  \approx 0.116                                                                                       \m . 
\ee 
In terms of this, 
\be 
\mbox{Prob}(\mS^{\sharp})   \es   \frac{3}{4} - \chi  
                         \approx  0.402                                                                   \m , 
\ee 
\be 
\mbox{Prob}(\mL^{\flat})   \es   1 - 7 \, \chi  
                         \approx  0.188                                                                   \m , 
\ee
\be 
\mbox{Prob}(\mL^{\sharp})   \es   7 \, \chi - \frac{3}{4}  
                         \approx  0.062                                                                   \m .   
\ee
Thirdly, 
\be
\mbox{Prob}(\mu^{\sharp}) = \left(\mbox{arc } > \alpha \, | \, \mbox{uniform} \right)     =    \frac{\delta}{3}
                                                                   \approx  0.155 
\ee 
and 
\be 
\mbox{Prob}\left(\frac{4 \, \pi}{3} < \mbox{arc } < \alpha \, | \, \mbox{uniform} \right)    =     \frac{1 - \delta}{3} 
                                                                        \approx  0.177                   \m . 
\ee 

\vspace{10in}

\section{Conclusion}

\subsection{The main results}

\n We considered the Isometric Shape-and-Scale Theory of $N$ points-or-particles on the carrier space $\mathbb{S}^1$. 
The constellation spaces for this are 
\be
\FrQ(N, \mathbb{S}^1) = \bigtimes_{i = 1}^N\mathbb{S}^1 = \mathbb{T}^N \m .
\ee  
The isometry group considered is $Isom(\mathbb{S}^1) = U(1)$, which, being topologically $\mathbb{S}^1$ again, 
leads to the shape-and-scale spaces thmselves being just 
\be
\FrR(N, \mathbb{S}^1) = \bigtimes_{i = 1}^n \mathbb{S}^1 = \mathbb{T}^n
\ee 
for 
$n := N - 1$. 
These shapes-and-spaces are productively described by Jacobi-like relative angle cluster coordinates for $N \geq 3$ (for $N = 2$ one relative angle will do).
Shape spaces are moreover not meaningful for this carrier space, 
due to topological identification globally invalidating the dilational generator's status as a generalized (similarity) Killing vector.

\m 

\n The minimal relationally nontrivial unit occurs for $N = 3$, providing the main focus of study of the current article; the corresponding shape-and-scale space is $\mathbb{T}^2$.  
This admits a natural identified parallelogram presentation.  
Coincidence-or-collision topological considerations equip this with a 2-tile tessellation. 

\m 

\n On the one hand, symmetry considerations refine this to a 12-tile tessellation. 
The latter suffices to understand mirror image and point-or-particle label identifications. 
This provides a lattice of 8 further shape-or-scale spaces, in particular the single-tile Leibniz space, 
which takes the form of the `30$^{\so}$-60$^{\so}$-90$^{\so}$ set square' triangle.  
The various discretely identified shape-or-scale spaces are topologically $\mathbb{S}^2$, a real projective space $\mathbb{RP}^2$ or discs $\mathbb{D}^2$.    

\m 

\n On the other hand, extending the topological features with the antipodal structure provides an 8-tile tessellation. 
The combined topological, symmetry and antipodal features provide a 24-tile tessellation, constituting the Lagrangian structure of the model. 
In particular, the partial antipodality lines split the Leibniz space tile into `short arc' (less than antipodal) and `long arc' configurations. 
This furnishes a shape-theoretic analogy with triangles, with partial antipodality analogous to rightness, and short and long arcs in analogy with acute and obtuse triangles. 
The equilateral triangle corresponds to $\Leib_{\sFrR}(3, \mathbb{S}^1$)'s most uniform state U.  

\m 

\n This analogy is further substantiated in shape space by the partial antipodality line meeting the edge opposite to it -- the shorter-arc uniformity line -- at right angles. 
While there is a corresponding topological analogy between geodesics of regularity in these two models, the triangle case of this makes a further right angle 
with its opposite side -- the collinearities -- while the $(3, \mathbb{S}^1)$ case meets its opposite side -- the longer-arc uniformity line -- obliquely. 
This renders the Jacobi structure of $(3, \mathbb{S}^1)$ more complicated than the triangle's. 
An expansion on previous articles' consideration of boundary conditions for geodesics colliding with Leibniz space edges is warranted, 
and is shown to partly depend on the form of a more extended patch of the tessellation by topology-and-symmetry notions.  

\m 

\n We finally provided a counterpart of Lewis Carroll's Pillow Problem \cite{Carroll, Guy-Portnoy} of what is Prob(obtuse) for triangles, 
which admits a shape-theoretic solution \cite{Small, MIT, A-Pillow} as well as further exactly calculable problems concerning restricting to isoscelessness and/or 
interplay with regularity \cite{A-Pillow}.  
For $(3, \mathbb{S}^1)$ these analogues are Prob(short arc), its restriction to configurations with one element of uniformity, and its interplay with regularity. 
These are slightly simpler to compute for $\mathbb{S}^1$ (flat-space triangles versus spherical triangles).  

\m

\n 3 points-or-particles on $\mathbb{S}^1$ is moreover very different from the $\mathbb{R}$ case, not least through having the above analogies with the $\mathbb{R}^2$ case.  
Comparing $\mathbb{S}^1$ and $\mathbb{R}$ carrier space cases for 3 points-or-particles is thereby sufficient to indicate that 
idea that quotienting out geometrical automorphisms banishes an incipient notion of absolute space is dead.  
Such indirect modelling is, rather, well capable of remembering the incipient absolute space's topology.
Thus topological considerations of Background Independence have become indispensible even in mechanics models. 
This motivates a) further comparison between relational theories on topologically distinct carrier spaces and b) comparison with how variable spatial topology is handled 
in generalizations of General Relativity (GR), which we address in the last two subsections below.

\subsection{Comparison with other carrier space models}

\n Since $\mathbb{S}^1$ is the first of the $\mathbb{T}^d$ and of the $\mathbb{RP}^d$ spaces as well as of the $\mathbb{S}^d$ ones, 
it is interesting to compare the current article's results with the Relational Theory on each of the first distinct torus, real projective space and sphere carrier spaces.  
We find that $\FrR(N, \mathbb{S}^1)$ has a single stratum, a key feature perpetuated by $\mathbb{T}^d$ carrier spaces but not 
$\mathbb{R}^d$, $\mathbb{S}^d$ or $\mathbb{RP}^d$ ones; see \cite{ATorus} for further details about $\mathbb{T}^d$ Shape-and-Scale Theory.  

\m 

\n See \cite{A-Monopoles} and \cite{ASphe} for more details of the corresponding  examples.  
In particular, maximal coincidences-or-collisions are separate strata for both $\mathbb{S}^2$ and $\mathbb{R}^3$ \cite{A-Monopoles}. 
The first of these furthermore possesses an intermediary stratum consisting of collinear shapes while the latter's stratum also contains totally antipodal shapes.
[I.e.\ shapes in which all the points-or-particles are distributed over a single point and its antipode.]  

\m 

\n Via the Relational Aufbau Principle, $\mathbb{S}^1$ Shape-and-Scale Theory is moreover crucial for developing 
$\mathbb{S}^d$ and $\mathbb{RP}^d$ shape-and-scale theories, occurring in particular in the analysis of collinear configurations on these other carrier spaces.

\subsection{Comparison with GR}

\n One way of removing the topological-level imprint of absolute space from one's model is to consider all the topologies at once  
(perhaps within some subclass: see below for some examples).  

\m 

\n At the level of configuration spaces, this involves 
\be
\Big\mbox{-}\FrQ = \coprod_{\tau \in \sFrT} \FrQ(\tau)  \mbox{ } .  
\ee
Such spaces might additionally be accorded further levels of mathematical structure, such as their own topology.  

\m 

\n At the level of actions, quantum operators, quantum path integrals, notions of information...  
this subsequently involves using corresponding `sums over topologies' to remove the effect of the element of choice of a particular topology.
Schematically, one replaces $\tau$-dependent objects $\FrO(\tau)$ by 
\be
\mbox{\Large S}_{\tau \in \sFrT} \FrO(\tau)                           \m .
\ee
\n{\bf Example 1} Let $\FrT$ be the set of all topological manifolds $\FrT$ (rather than topological spaces more generally, and taken to exclude cases `with boundary' and so on).  
Further common class restrictions include the following. 

\m  

\n 1) Considering only the connected manifolds \cite{GH92}.  

\m 

\n 2) Considering only the compact manifolds \cite{Battelle}.

\m 

\n 3) Considering only the orientable manifolds.

\m 

\n 4) Considering only topological manifolds of a fixed dimension.

\m 

\n 5) Considering only a dimensional series of manifolds, e.g. $\mathbb{R}^d$, $\mathbb{T}^d$, $\mathbb{S}^d$ or $\mathbb{RP}^d$.

\m 

\n In the context of point-or-particle models, variable $N$ can also be considered to have topological content \cite{ABook}
(at the level of allowing point-or-particle fission and fusion and/or particle creation and annihilation).
Because of this, 
\be
\coprod_{N \in \mathbb{N}_0} \m \mbox{ and } \m \sum_{N \in \mathbb{N}_0} 
\ee
operations are also an option among the various ingredients considered.  

\m 

\n One possibility is then 
\be
\Big\mbox{-}\FrR(d)  \es  \coprod_{\sFrC^d \in \sFrM(\mbox{\scriptsize connected})} \, \coprod_{N \in \mathbb{N}_0} \, \FrR(N, \FrC^d) 
\ee 
for a given $d$.  

\m 

\n For $d = 1$, this returns 
\be 
\Big\mbox{-}\FrR(1)  \es  \coprod_{\sFrC^1 = \mathbb{R}, \mathbb{S}^1} \, \coprod_{N \in \mathbb{N}_0} \, \FrR(N, \FrC^1) \mbox{ } . 
\label{Big-1}
\ee 
For $d = 2$, assuming additionally $\FrC^2$ is compact and oriented, 
\be 
\Big\mbox{-}\FrR(2)  \es  \coprod_{\sg = 0}^{\infty} \, \coprod_{N \in \mathbb{N}_0} \, \FrR(N, \FrC^2_{\sg}) \mbox{ } ,  
\label{Big-2}
\ee 
where g is the genus.
Dropping the oriented condition just adds a 
\be 
\coprod_{o = \pm 1}
\ee 
to the string of disjoint unions and a corresponding o-index to the carrier space.  

\m

\n For $d \geq 3$, disjoint union of, and `sum over', all topologies acquires a formal character. 

\m  

\n Finally one has 
\be 
\Grand\mbox{-}\FrR  \es  \coprod_{d = 1}^{\infty} \,  \coprod_{\sFrC^d \in \sFrM(\mbox{\scriptsize connected)}} \, \coprod_{N \in \mathbb{N}_0} \, \FrR(N, \FrC^d)  \m .
\ee 
\n GR counterparts of point-or-particle relational spaces are well-known. 

\m 

\n 0) for a given 3-$d$ spatial topology $\bupSigma_3$, 
\be 
\Riem(\bupSigma_3)
\ee 
is the space of all positive-definite 3-metrics on $\bupSigma_3$.  
This is GR's analogue of constellation space. 

\m 

\n 1) The group of physically irrelevant automorphisms considered is now $Diff(\bupSigma_3)$. 

\m 

\n 2) The corresponding quotient is Wheeler's \cite{Battelle} 
\be 
\Superspace(\bupSigma_3) \es \frac{\Riem(\bupSigma_3)}{Diff(\bupSigma_3)}                       \m ,
\ee  
further studied in \cite{DeWitt67, DeWitt, Fischer, FM96, Giu09}. 
A particular feature that this shares with relational spaces is that it is stratified; it is now 3-metrics possessing Killing vectors that constitute the nontrivial strata. 

\m 

\n 3) Fischer \cite{Fischer} additionally entertatined the concept of a 
\be 
\Big\mbox{-}\Superspace(\bupSigma_3) \es \coprod_{\bupSigma_3 \in \sFrT^3} \, \Superspace(\bupSigma_3)  \m . 
\ee 
This would usually be considered for $\sFrT^3$ additionally connected, compact and orientable.
It is a model underlying a variant of GR which furthermore allows for spatial topology change; by technical necessity, it remains a merely formal model.  
With $d$-dimensional superspace being as straightforward to define, one can additionally conceive of 
\be 
\Grand\mbox{-}\Superspace  \es \coprod_{d = 1}^\infty \, \coprod_{\bupSigma_d \in \sFrT^d} \, \Superspace(\bupSigma_d)  \m ,  
\ee 
possibly suppressing $d = 1$ and 2 contributions since these have no degrees of freedom.  
This is now for a variant of GR which allows for topology change including change of spatial dimension.  

\m 

\n In this light, one can view (\ref{Big-1}) and (\ref{Big-2}) (with or without variable-$N$) as models of Fischer's Big Superspace. 
Among these, (\ref{Big-1}) is particularly simple, and requires the current article's development of $\mathbb{S}^1$ relationalism 
alongside its more established $\mathbb{R}$ counterpart \cite{I, II}. 
Moreover, as we shall see in \cite{Top-Shapes}, the rubber configurations allow for a calculable analogue of even Grand Superspace. 
Here again $\mathbb{S}^1$ plays a key role, as one of only three connected manifold without boundary rubber carrier space classes, 
the others being $\mathbb{R}$ and the arbitrary such carrier space with dimension $\geq 2$.  

\m 

\n{\bf Acknowledgments} I thank Chris Isham and Don Page for discussions about configuration space topology, geometry, quantization and background independence. 
I also thank Jeremy Butterfield and Christopher Small for encouragement. 
I thank Don, Jeremy, Enrique Alvarez, Reza Tavakol and Malcolm MacCallum for support with my career. 
Chris Isham's Review \cite{I84} on the role of topology in quantization was particularly inspirational as regards the current article.   

\vspace{10in}

\begin{appendices}

\section{(3, 1) Example}

The configurations for 3 particles on the line $\mathbb{R}$ are in Fig \ref{(3,1)-Configs}, with the corresponding lattice of discrete quotients in Fig \ref{R(3,1)-Metric-Lattice}.
The main point of this Appendix is that these are much simpler than their $\mathbb{S}^1$ counterparts. 
%
{            \begin{figure}[!ht]
\centering
\includegraphics[width=0.8\textwidth]{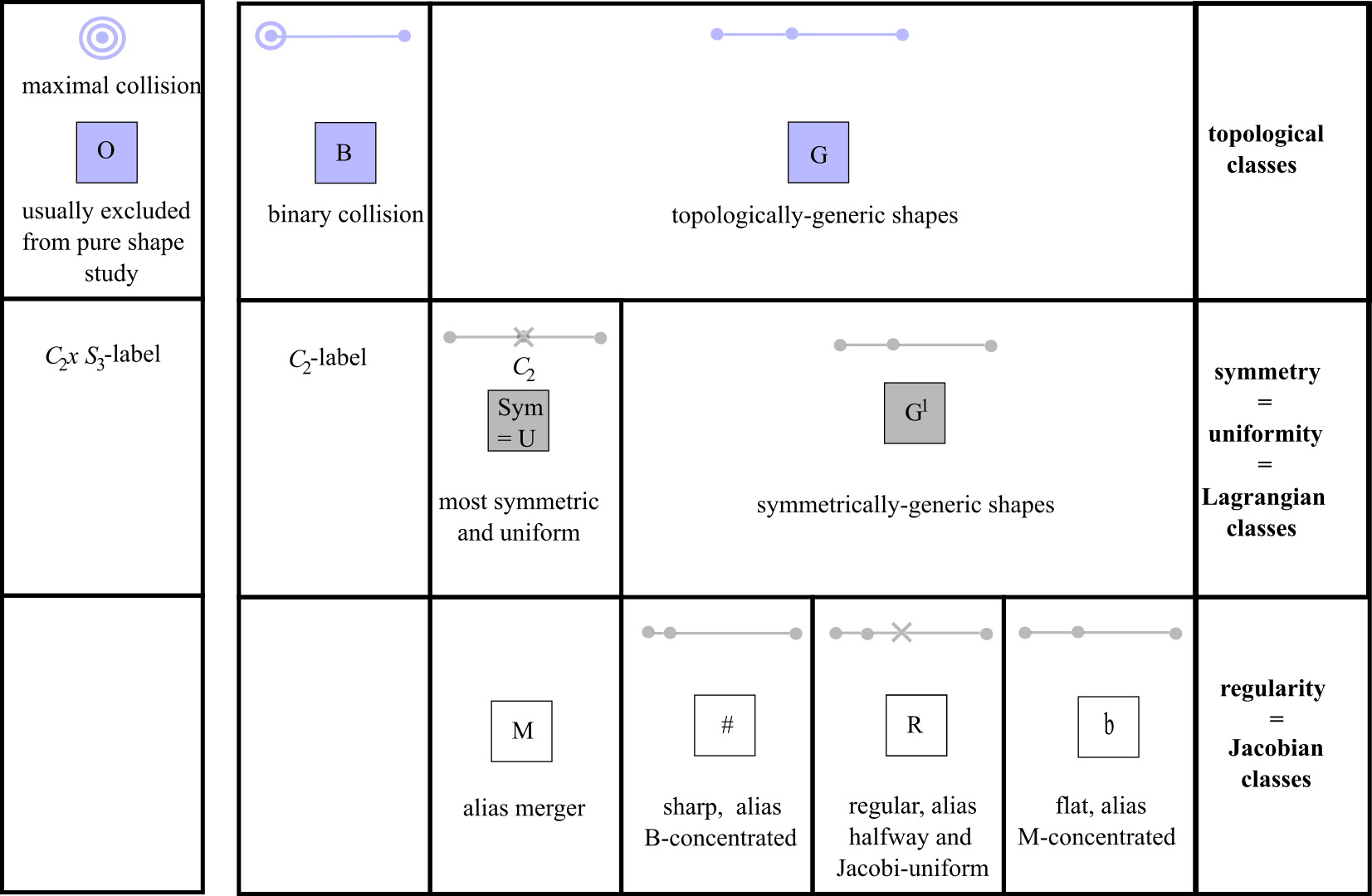}
\caption[Text der im Bilderverzeichnis auftaucht]{        \footnotesize{Types of configuration for $N = 3$ points-or-particles on $\mathbb{R}$ at the topology, topology-and-symmetry 
(= Lagrangian for this model) and Jacobian (= regularity for this model) level of structure.
Note in particular that there are far less classes than for the current article's main model of $N = 3$ points-or-particles on $\mathbb{S}^1$ instead.    
} }
\label{(3,1)-Configs} \end{figure}          }
%
{            \begin{figure}[!ht]
\centering
\includegraphics[width=0.45\textwidth]{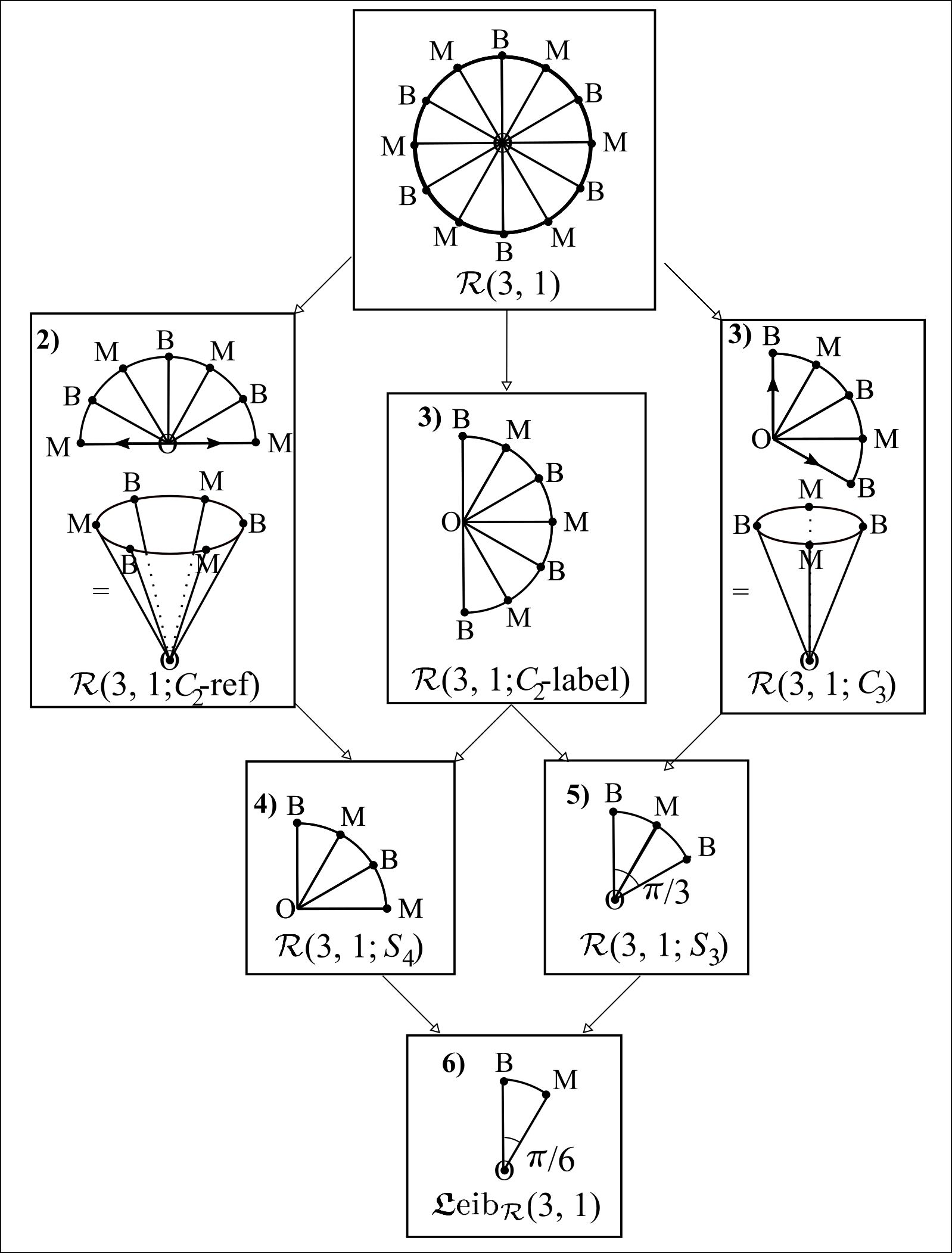}
\caption[Text der im Bilderverzeichnis auftaucht]{        \footnotesize{Lattice of discrete quotients for (3, 1).} }
\label{R(3,1)-Metric-Lattice} \end{figure}          }

\vspace{10in}

\section{(3, 2) Example}
%
{            \begin{figure}[!ht]
\centering
\includegraphics[width=0.60\textwidth]{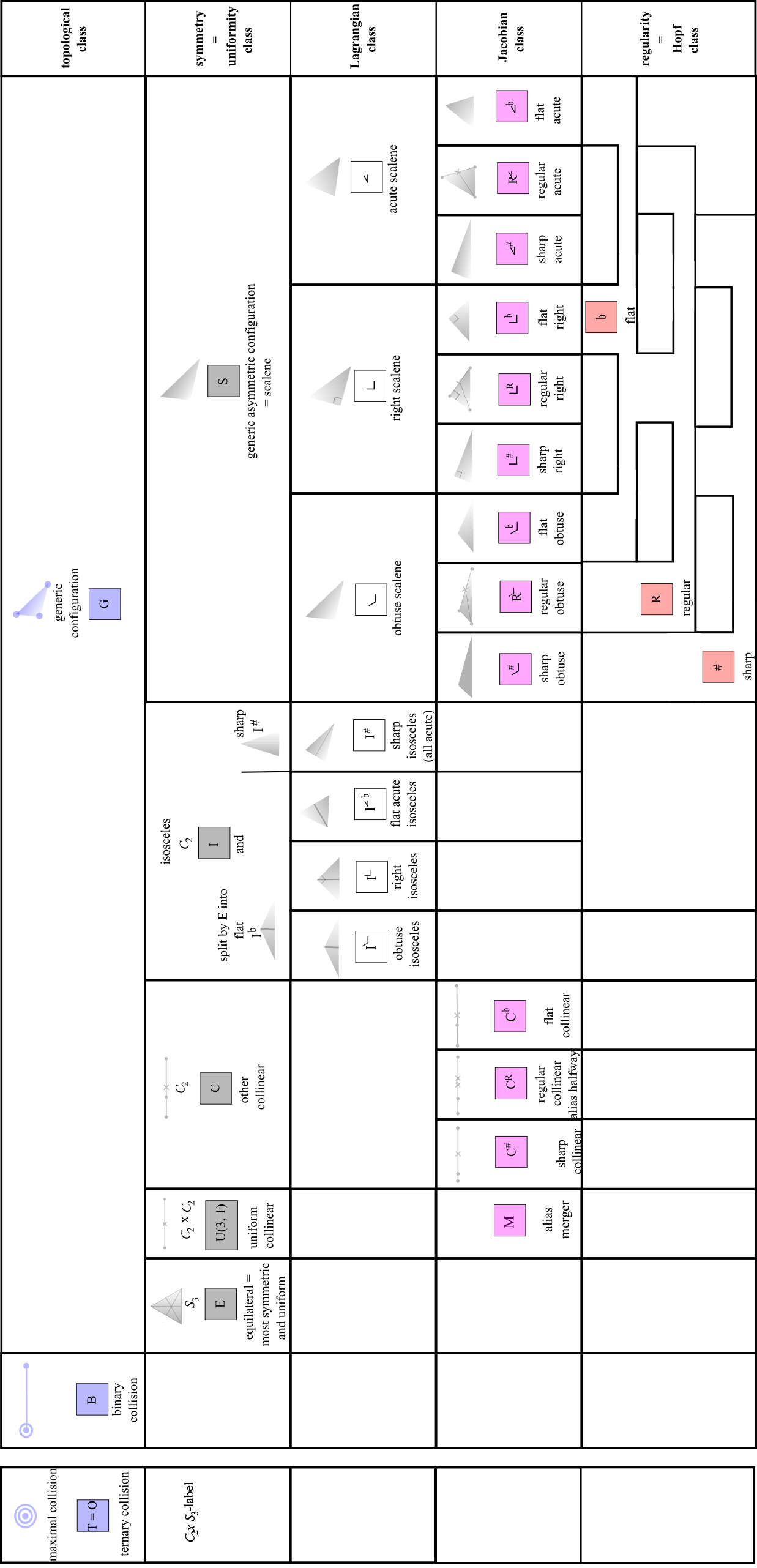}
\caption[Text der im Bilderverzeichnis auftaucht]{        \footnotesize{The topology, 
                                                                            symmetry, 
																			uniformity, 
																			Lagrangian, 
																			Jacobian and 
																			regularity classes for triangular shapes in $\mathbb{R}^2$; 
																		for triangles, the last of these is moreover also the Hopf class \cite{III}. } }
\label{Triangle-Configs} \end{figure}          }

\vspace{10in}

\n The configurations for triangles -- 3 points-or-particles in 2-$d$ -- $\mathbb{R}$ are in Fig \ref{Triangle-Configs}, 
with the corresponding Leibniz spaces in Figs \ref{Leib(3, 2)-Sym}--\ref{Leib(3, 2)-Hopf} to various levels of structure.  
The main point of this Appendix is the analogies between these and the article's $(3, \mathbb{S}^1)$ counterparts, each of which is marked on the figures' captions.  
%
{            \begin{figure}[!ht]
\centering
\includegraphics[width=0.85\textwidth]{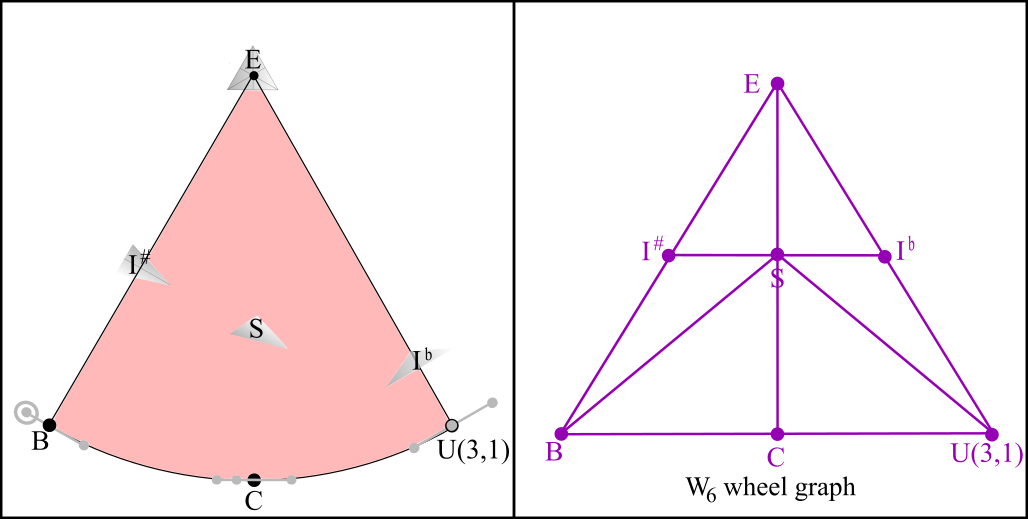}
\caption[Text der im Bilderverzeichnis auftaucht]{        \footnotesize{a) $(\Sym \bigcup \Top)(\Leib_{\tFrS}(3, 2)) = (\Uni \bigcup \Top)(\Leib_{\tFrS}(3, 2)))$: 
the Leibniz space for triangular shapes in $\mathbb{R}^2$ at the level of symmetry as well as topology, or, alternatively, of (Lagrangian) uniformity instead of symmetry.   

\m

\n b) The corresponding adjacency graph, $\FrA((\Sym \bigcup \Top)(\Leib_{\tFrS}(3, 2)))$ 
At the level of unlabelled graphs, this is the 6-spoked wheel graph $W_6$ and thus isomorphic to the current article's $\FrA(\Sym(\Leib_{\tFrR}(3, \mathbb{S}^1)))$.
This gives the analogies of collinearity C and isoscelesness I as edge concepts to B and configurations with one element of uniformity u, 
						 and equilaterality E, B and U(3, 1) as vertices in some order to O, total antipodality A and the most uniform configuration U(3, $\mathbb{S}^1$).
} }
\label{Leib(3, 2)-Sym} \end{figure}          }
%
{            \begin{figure}[!ht]
\centering
\includegraphics[width=0.85\textwidth]{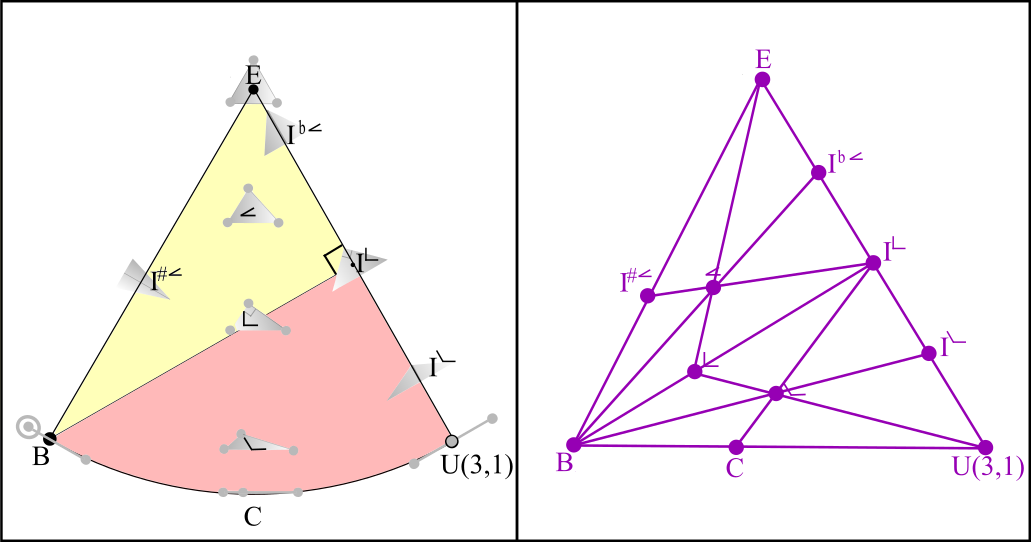}
\caption[Text der im Bilderverzeichnis auftaucht]{        \footnotesize{a) $\Lag(\Leib_{\tFrS}(3, 2))$: 
the Leibniz space for triangular shapes in $\mathbb{R}^2$ at the Lagrangian level, which amounts to further including the rightness structure.  

\m 

\n b) The corresponding adjacency graph, $\FrA(\Lag(\Leib_{\tFrS}(3,  2)))$ 
At the level of unlabelled graphs, this is two 6-spoked wheel graphs $W_6$ joined along two adjacent-perimeter edges, 
by which it is isomorphic to the current article's $\FrA(\Lag(\Leib_{\tFrR}(3, \mathbb{S}^1)))$.
This gives the further analogies B to A specifically, 
and rightness $\perp$ to partial antipodality A2, splitting regions of acuteness and obtuseness corresponding to short arcs to long arcs respectively.
Furthermore, sharp isoscelesness $\mI^{\sharp}$ corresponds to B, flat isoscelessness $\mI^{\flat}$ to $\muu^{\sS}$, and collinearity to $\muu^{\sL}$.  
} }
\label{Leib(3, 2)-Lag} \end{figure}          }

\vspace{10in}

{            \begin{figure}[!ht]
\centering
\includegraphics[width=1.0\textwidth]{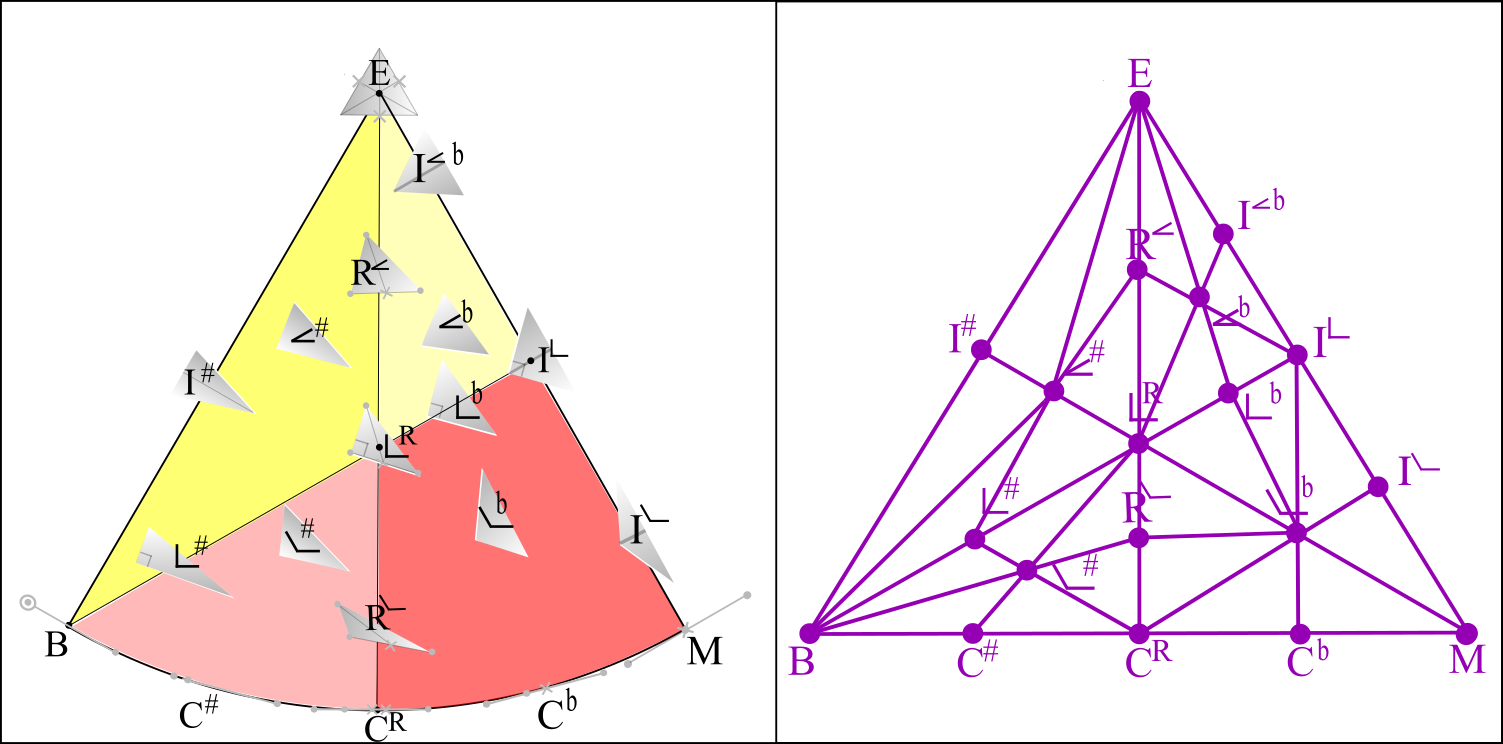}
\caption[Text der im Bilderverzeichnis auftaucht]{        \footnotesize{a) $\Jac(\Leib_{\tFrS}(3, 2)$:  
the Leibniz space for triangular shapes in $\mathbb{R}^2$ at the Jacobian level structure, which amounts to further including the regularity structure.  

\m 

\n b)  The corresponding adjacency graph $\FrA(\Jac(\Leib_{\tFrS}(3, 2)))$.  
This is drawn to exhibit its unlabelled-graph-level isomorphism with the current article's $\FrA(\Jac(\Leib_{\tFrR}(3, \mathbb{S}^1)))$, 
with the regular right triangle now playing the role of central vertex.}}
\label{Leib(3, 2)-Full-Jac} \end{figure}          }
%
{            \begin{figure}[!ht]
\centering
\includegraphics[width=0.85\textwidth]{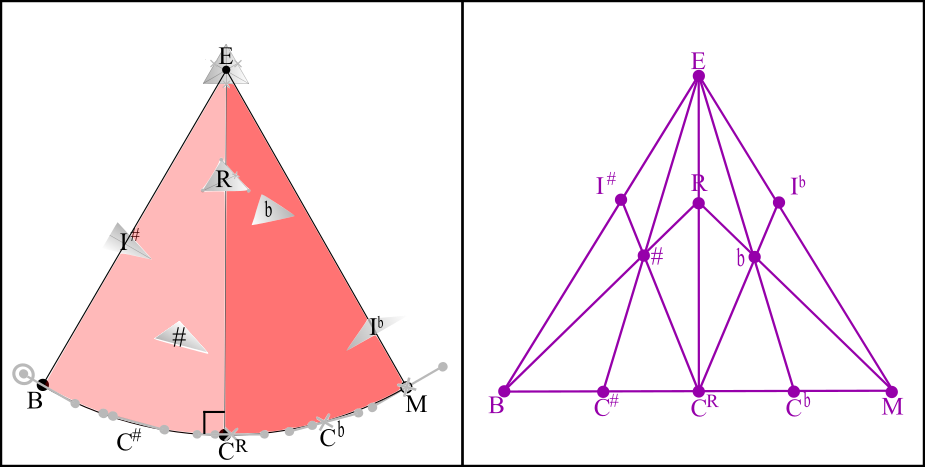}
\caption[Text der im Bilderverzeichnis auftaucht]{        \footnotesize{a) $\Reg(\Leib_{\tFrS}(3, 2))$: 
the Leibniz space for triangular shapes in $\mathbb{R}^2$ including the regularity structure but not the rightness structure.
In this case, this can also be envisaged as combining the topological structure with {\sl all} the uniformity structure. 
It furthermore coincides with the Hopf structure special to triangles from their shape space (portions of) sphere additionally realizing the Hopf bundle.  

\m

\n b) The corresponding adjacency graph $\FrA(\Reg(\Leib_{\tFrS}(3, 2))) = \FrA(\Hopf(\Leib_{\tFrS}(3, 2)))$; 
this is clearly yet another instance of two $\mW_6$ glued along two adjacent perimeter edges, rendering it isomorphic at the level of unlabelled graphs with 
Figs \ref{Leib(3, 2)-Lag}.b), \ref{Leib(3,S1)-Lag}.b) and in particular with the current article's regularity structure of Fig \ref{Leib(3,S1)-Reg}.b). 
The analogy here is that the two incipient notions of regularity line up at the topological level. 
However, at the metric level, regularity impacts collinearity at right angles, 
whereas regularity impacts uniformity at a different action causing reflective prolongation.  
So the preceding Figure is {\sl all} the Jacobian structure, unlike Fig \ref{Leib(3,S1)-Reg}.  
} }
\label{Leib(3, 2)-Hopf} \end{figure}          }
    
\vspace{10in}

\end{appendices}


\end{document}